\pdfoutput=1
\documentclass[11pt,twoside,a4paper,cmspaper,final,collab]{cms-tdr}
\def\svnVersion{174b2d9}\def\svnDate{2026/08/11}\def\cmsCernNoTag{CERN-EP-2026-091}\def\cmsCernDate{\today}\def\cmsMessage{Published in the Journal of High Energy Physics as \href{http://dx.doi.org/10.1007/JHEP08(2026)019}{\doi{10.1007/JHEP08(2026)019}.}}
\begin{document}\cmsNoteHeader{HIG-25-018}

\newlength\cmsTabSkip\setlength{\cmsTabSkip}{1ex}

\newcommand{\pp}{\ensuremath{\Pp\Pp}\xspace}
\newcommand{\ggF}{\ensuremath{\Pg\Pg\text{F}}\xspace}
\newcommand{\VBF}{\ensuremath{\text{VBF}}\xspace}
\newcommand{\HH}{\ensuremath{\PH\PH}\xspace}
\newcommand{\bbWW}{\ensuremath{\bbbar\PW\PW}\xspace}
\newcommand{\Htobb}{\ensuremath{\PH\to\bbbar}\xspace}
\newcommand{\DY}{\ensuremath{\text{DY}}\xspace}
\newcommand{\DYHF}{\ensuremath{\text{DY}+\text{HF}}\xspace}
\newcommand{\DYLF}{\ensuremath{\text{DY}+\text{LF}}\xspace}
\newcommand{\Wjets}{\ensuremath{\PW+\text{jets}}\xspace}
\newcommand{\ttV}{\ensuremath{\ttbar\PV}\xspace}
\newcommand{\ttVV}{\ensuremath{\ttbar\PV\PV}\xspace}
\newcommand{\tttt}{\ensuremath{\ttbar\ttbar}\xspace}
\newcommand{\diboson}{\ensuremath{\PV\PV}\xspace}
\newcommand{\triboson}{\ensuremath{\PV\PV\PV}\xspace}
\newcommand{\singlet}{\text{single-\PQt/\PAQt}\xspace}
\newcommand{\kappaL}{\ensuremath{\kappa_{\lambda}}\xspace}
\newcommand{\kappaVV}{\ensuremath{\kappa_{2\PV}}\xspace}
\newcommand{\mll}{\ensuremath{m_{\Pell\Pell}}\xspace}
\newcommand{\ptll}{\ensuremath{\pt^{\Pell\Pell}}\xspace}
\newcommand{\ptee}{\ensuremath{\pt^{\Pe\Pe}}\xspace}
\newcommand{\ptmumu}{\ensuremath{\pt^{\PGm\PGm}}\xspace}
\newcommand{\Njets}{\ensuremath{N_{\text{jets}}}\xspace}

\newcommand{\muR}{\ensuremath{\mu_{\text{R}}}\xspace}
\newcommand{\muF}{\ensuremath{\mu_{\text{F}}}\xspace}
\newcommand{\muRF}{\ensuremath{\mu_{\text{R/F}}}\xspace}

\newcommand{\mumu}{\ensuremath{\PGmp\PGmm}\xspace}
\newcommand{\ee}{\ensuremath{\Pep\Pem}\xspace}
\newcommand{\emu}{\ensuremath{\Pepm\PGmmp}\xspace}

\newcommand{\NNmult}{\ensuremath{\text{NN}_{\text{cat}}}\xspace}
\newcommand{\NNggF}{\ensuremath{\text{NN}_{\text{ggF}}}\xspace}
\newcommand{\NNVBF}{\ensuremath{\text{NN}_{\text{VBF}}}\xspace}
\newcommand{\ParticleNet}{{\scshape ParticleNet}\xspace}

\newcommand{\catb}{\ensuremath{\text{1\PQb}}\xspace}
\newcommand{\catbb}{\ensuremath{\text{$\geq$2\PQb}}\xspace}
\newcommand{\catboosted}{\ensuremath{\text{boosted}}\xspace}
\newcommand{\SRggF}{\ensuremath{\text{SR}_{\ggF}}\xspace}
\newcommand{\SRggFb}{\ensuremath{\text{SR}_{\ggF}^{\catb}}\xspace}
\newcommand{\SRggFbb}{\ensuremath{\text{SR}_{\ggF}^{\catbb}}\xspace}
\newcommand{\SRggFboosted}{\ensuremath{\text{SR}_{\ggF}^{\catboosted}}\xspace}
\newcommand{\SRVBF}{\ensuremath{\text{SR}_{\VBF}}\xspace}
\newcommand{\SRVBFb}{\ensuremath{\text{SR}_{\VBF}^{\catb}}\xspace}
\newcommand{\SRVBFbb}{\ensuremath{\text{SR}_{\VBF}^{\catbb}}\xspace}
\newcommand{\SRVBFboosted}{\ensuremath{\text{SR}_{\VBF}^{\catboosted}}\xspace}

\newcommand{\fDYcorr}{\ensuremath{f^{\text{corr}}_{\DY}}\xspace}

\newcommand{\lumitot}{\ensuremath{62\fbinv}\xspace}
\newcommand{\resultGoF}{\ensuremath{0.19}\xspace}
\newcommand{\normTTres}{\ensuremath{0.95}\xspace}
\newcommand{\normTTresUp}{\ensuremath{0.03}\xspace}
\newcommand{\normTTresDn}{\ensuremath{0.02}\xspace}
\newcommand{\normTTboost}{\ensuremath{0.82}\xspace}
\newcommand{\normTTboostUp}{\ensuremath{0.08}\xspace}
\newcommand{\normTTboostDn}{\ensuremath{0.07}\xspace}
\newcommand{\resultLimitHHexp}{\ensuremath{18.5}\xspace}
\newcommand{\resultLimitHHobs}{\ensuremath{12.0}\xspace}
\newcommand{\resultKappaL}{\ensuremath{3.4}\xspace}
\newcommand{\resultKappaLMinObs}{\ensuremath{-9.1}\xspace}
\newcommand{\resultKappaLMaxObs}{\ensuremath{15.7}\xspace}
\newcommand{\resultKappaLMinExp}{\ensuremath{-13.3}\xspace}
\newcommand{\resultKappaLMaxExp}{\ensuremath{19.8}\xspace}
\newcommand{\resultKappaVVMinObs}{\ensuremath{-0.20}\xspace}
\newcommand{\resultKappaVVMaxObs}{\ensuremath{2.25}\xspace}
\newcommand{\resultKappaVVMinExp}{\ensuremath{-0.48}\xspace}
\newcommand{\resultKappaVVMaxExp}{\ensuremath{2.54}\xspace}
\newcommand{\resultCombinationLimitHHexp}{\ensuremath{12.7}\xspace}
\newcommand{\resultCombinationLimitHHobs}{\ensuremath{7.9}\xspace}

\cmsNoteHeader{HIG-25-018}
\title{Search for Higgs boson pair production in the \bbWW decay channel with two leptons in the final state using proton-proton collision data at $\sqrt{s}=13.6\TeV$}
\date{\today}

\abstract{
A search for Higgs boson pair production is presented, targeting final
states where one Higgs boson decays to a bottom quark-antiquark pair and the other Higgs boson decays to two \PW bosons, both of which decay leptonically, to an electron or a muon, and a neutrino. For the first time, the search is conducted with proton-proton collision data from the LHC at $\sqrt{s}=13.6\TeV$, recorded with the CMS detector in 2022 and 2023 and corresponding to an integrated luminosity of \lumitot. The results are consistent with the standard model predictions. An upper limit of \resultLimitHHobs times the standard model prediction at 95\% confidence level is set on the Higgs boson pair production cross section, with an expected limit of \resultLimitHHexp. The results are also used to constrain the strength of the trilinear self-coupling of the Higgs boson, as well as of the quartic coupling between two Higgs bosons and two vector bosons.
}

\hypersetup{%
pdfauthor={CMS Collaboration},%
pdftitle={Search for Higgs boson pair production in the bbWW decay channel with two leptons in the final state using proton-proton collision data at sqrt(s) = 13.6 TeV},%
pdfsubject={CMS},%
pdfkeywords={CMS, HH}} %

\maketitle

\section{Introduction}
\begin{figure}[!htp]
  \centering
  \includegraphics[height=0.12\textheight]{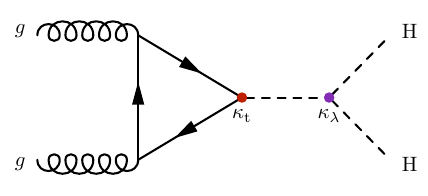}
  \includegraphics[height=0.12\textheight]{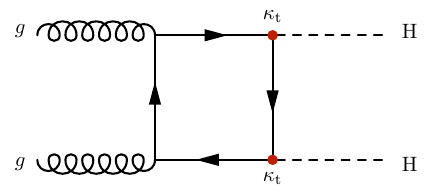}\\
  \includegraphics[height=0.15\textheight]{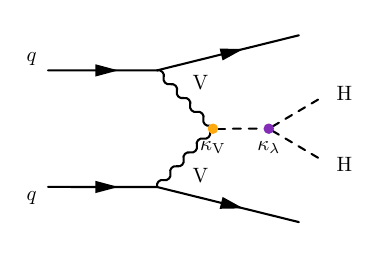}
  \includegraphics[height=0.15\textheight]{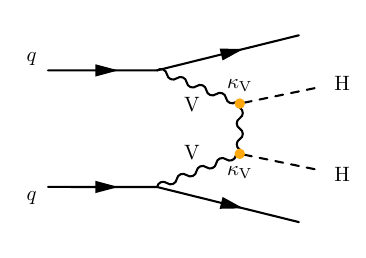}
  \includegraphics[height=0.15\textheight]{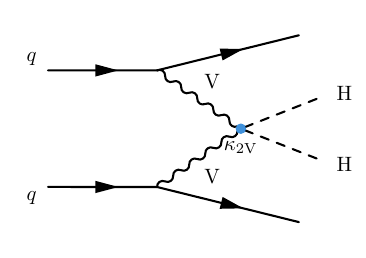}
  \caption{
    Leading-order Feynman diagrams of \HH production in the \ggF
    production mode assuming top quarks in the fermion loop (top row)
    and in the \VBF production mode (bottom row) in the SM.
  }
  \label{fig:intro:diagrams}
\end{figure}

The measured properties of the Higgs boson (\PH)
discovered~\cite{ATLAS:2012yve,CMS:2012qbp,CMS:2013btf} by the ATLAS
and CMS Collaborations at the CERN LHC are consistent with
the expectations of the standard model (SM) Brout--Englert--Higgs
(BEH) mechanism~\cite{Englert:1964et,Higgs:1964ia,Higgs:1964pj,Guralnik:1964eu,Higgs:1966ev,Kibble:1967sv,Weinberg:1967tq,Salam:1968rm} within the current experimental precision~\cite{ATLAS:2022vkf,CMS:2022dwd}.
One aspect of the BEH mechanism that is largely unknown
experimentally is the exact shape of the Higgs potential, which, in the SM,
is related to the trilinear self-coupling of the Higgs boson and
can be directly probed using Higgs boson pair (\HH) production~\cite{Heinrich:2019bkc}.
In the SM, the dominant \HH production mode at the LHC is gluon-gluon
fusion (\ggF), followed by vector boson fusion (\VBF) production, with
cross sections of
$34.1^{+2.2}_{-7.9}\unit{fb}$~\cite{Grazzini:2018bsd,Karlberg:2024zxx,Baglio:2020wgt}
and $1.87\pm0.05\unit{fb}$~\cite{Dreyer:2020xaj,Dreyer:2018qbw},
respectively, at the centre-of-mass energy ($\sqrt{s}$) of 13.6\TeV for a Higgs boson mass of 125\GeV.
The \VBF production mode is furthermore sensitive to the quartic
coupling between two Higgs bosons and two vector
bosons~\cite{Bishara:2016kjn}.
Representative Feynman diagrams of \HH production in the \ggF and \VBF
modes are shown in Fig.~\ref{fig:intro:diagrams}.

The \HH production process has not yet been observed.
Previous searches for \HH production have been performed by the ATLAS
and CMS Collaborations with LHC Run 2 data at 13\TeV in various
final states and their
combinations~\cite{ATLAS:2024ish,CMS:2025ngq,CMS:2026nuu}.
Recently, the ATLAS Collaboration also published a search for \HH
production in the final state with a bottom quark-antiquark pair (\bbbar) and two photons,
combining both LHC Run 2 and Run 3 data at
13\TeV and 13.6\TeV, respectively~\cite{ATLAS:2025hhd}.

This article presents a search for non-resonant \HH production,
targeting final states where one Higgs boson decays to \bbbar and the
other Higgs boson decays to two \PW bosons, both of which decay leptonically to an electron or a muon, and a neutrino.
Previous searches in the \bbWW channel with the Run 2 data set were
performed by the CMS Collaboration in final states with one or two
leptons, setting an observed (expected) upper limit on the inclusive
\HH production cross section of 14 (18) times the SM prediction at
95\% confidence level (\CL)~\cite{CMS:2024rgy}, and by the ATLAS Collaboration in the
two-lepton final state with an upper limit of 9.7 (16.2) times the SM
prediction at 95\% \CL~\cite{ATLAS:2023elc}.

In this search, events are selected with exactly two oppositely charged leptons (\mumu, \ee, \emu).
The selected events are categorised based on the output of a
multiclassification neural network (NN) into several regions targeting
\HH signal events from \ggF or \VBF production, or background events
from different processes.
The categorisation is further refined based on the \PQb-tagged jet
multiplicity and the presence of a jet from a Lorentz-boosted \Htobb candidate.
The events in the signal-enriched regions are then classified by dedicated binary NNs designed to separate ggF or VBF signal events from background events.
The output distributions of the binary NNs, together with the
event yield in the background-enriched regions, enter the final fit as sensitive observables.
An upper limit is set on the inclusive \HH production cross section,
and the result is used to constrain the strength of the trilinear
Higgs boson self-coupling as
well as the quartic coupling between two Higgs bosons and two vector
bosons.
Throughout this article, the values of the Higgs boson couplings are
expressed in terms of coupling modifiers $\kappa$, denoting the coupling
strength relative to the SM prediction~\cite{Heinemeyer:2013tqa}.
The modifier of the trilinear coupling is denoted by \kappaL, and that of the quartic
coupling by \kappaVV.

Several refinements have been implemented relative to the previous search~\cite{CMS:2024rgy}.
A new classification strategy, utilising
a staged multiclassification-plus-binary NN approach, resulted in an
improvement in sensitivity of approximately 50\%, evaluated as the
exclusion limit expected with a dataset of the same size as used in
Ref.~\cite{CMS:2024rgy}.
Furthermore, an improved trigger strategy for \ee (\emu) events led to an increase of the signal yield in this channel by 5\% (1\%) inclusively and up to 17\% (3\%) in the most  signal-like bins of the final discriminant distribution.
Finally, an improved \PQb tagging algorithm was used,
and the analysis benefits from the
higher centre-of-mass energy of Run 3, resulting in a 5\% increase in
the ratio of signal events to the square root of the number of
background events.

This article is structured as follows.
The CMS detector as well as the object and event reconstruction are described in Section~\ref{sec:objects}, followed by a description of the data and the simulated event samples in Section~\ref{sec:samples}.
The event selection and the analysis strategy are detailed in Section~\ref{sec:strategy}, and systematic uncertainties are discussed in Section~\ref{sec:systematics}.
The statistical analysis and the results obtained are presented in Section~\ref{sec:results}.
Finally, a summary is provided in Section~\ref{sec:summary}.
Tabulated results are provided in the HEPData record for this
analysis~\cite{hepdata}.

\section{The CMS detector and event reconstruction}\label{sec:objects}
The CMS apparatus~\cite{CMS:2008xjf,CMS:2023gfb} is a multipurpose, nearly hermetic detector, designed to trigger on~\cite{CMS:2020cmk,CMS:2016ngn,CMS:2024aqx} and identify muons, electrons, photons, and (charged and neutral) hadrons~\cite{CMS:2020uim,CMS:2018rym,CMS:2014pgm}.
Its central feature is a superconducting solenoid of 6\unit{m} internal diameter, providing a magnetic field of 3.8\unit{T}.
Within the solenoid volume are a silicon pixel and strip tracker, a lead tungstate crystal electromagnetic calorimeter (ECAL), and a brass and scintillator hadron calorimeter, each composed of a barrel and two endcap sections.
Forward calorimeters extend the pseudorapidity coverage provided by the barrel and endcap detectors.
Muons are reconstructed using gas-ionisation detectors interleaved with the layers of the steel flux-return yoke outside the solenoid.
More detailed descriptions of the CMS detector, together with a definition of the coordinate system used and the relevant kinematic variables, can be found in Refs.~\cite{CMS:2008xjf,CMS:2023gfb}.
A global ``particle-flow'' (PF) algorithm~\cite{CMS:2017yfk} aims to reconstruct all individual particles in an event, with an optimised combination of information from the various elements of the CMS detector.
The reconstructed particles are used to build jets and missing transverse momentum (\ptmiss)~\cite{CMS:2016lmd,CMS:2019ctu}.

Events of interest are selected using a two-tiered trigger system.
The first level, composed of custom hardware processors, uses information from the calorimeters and muon detectors to select events at a rate of around 100\unit{kHz} within a fixed latency of 4\mus~\cite{CMS:2020cmk}.
The second level, known as the high-level trigger, consists of a farm of processors running a version of the full event reconstruction software optimised for fast processing, and reduces the event rate to a few kHz before data storage~\cite{CMS:2016ngn,CMS:2024aqx}.

The reconstruction of muons relies on a combination of measurements in the tracker and in the muon detectors~\cite{CMS:2018rym}.
The muons are identified based on the quality of the combined track fit and on the number of hits in the different tracking detectors, with an efficiency of about 90\%, including the isolation requirements described below.
The transverse momentum (\pt) resolution in the barrel is better than 1\% (7\%) for muons with \pt up to 100\GeV (1\TeV)~\cite{CMS:2018rym}.
Muons are required to lie within the acceptance of the muon system, covering pseudorapidity ($\eta$) values of $\abs{\eta}<2.4$, and to have $\pt>25$ and 15\GeV for the \pt-leading and subleading muons, respectively.

Electrons are reconstructed by combining the momentum measurement in the tracker with the energy measurement of the corresponding cluster in the ECAL and with the energy sum of all bremsstrahlung photons, obtained from the ECAL, spatially compatible with originating from the electron track~\cite{CMS:2020uim,CMS-DP-2020-021}.
The electrons are identified using criteria on the cluster shape in the ECAL, the track quality, and the compatibility between the tracker and ECAL measurements, corresponding to an identification efficiency of approximately 80\%, including the isolation requirements described below.
The momentum resolution for electrons with \pt of about 45\GeV ranges from 1.6 to 5\%~\cite{CMS:2020uim,CMS-DP-2020-021}.
Electrons are required to have $\abs{\eta}<2.5$ as well as $\pt>25$ and 15\GeV for the \pt-leading and subleading electrons, respectively.
Electrons reconstructed in the transition region $1.44<\abs{\eta}<1.56$ between the barrel and the endcap calorimeters are discarded.

Isolation requirements are imposed on the muons and electrons with the aim of suppressing leptons originating from decays of heavy hadrons inside jets.
The isolation is based on the scalar \pt sum of all PF objects in a cone of radius 0.3 in the $\eta$-$\phi$ plane around the track direction of the lepton, where $\phi$ denotes the azimuthal angle, and it is corrected by removing contributions arising from additional proton-proton (\pp) interactions in the same or an adjacent bunch crossing (pileup)~\cite{Sirunyan:2020foa}.

Hadronic jets are clustered from the PF objects using the anti-\kt
algorithm~\cite{Cacciari:2008gp,Cacciari:2011ma} with a distance
parameter of 0.4 (jets) and of 0.8 (large-radius jets).
No matching is performed between jets and large-radius jets, but they are treated as independent objects; thus, a large-radius jet will typically also appear as one or several corresponding jets in the event.
The pileup-per-particle identification algorithm (PUPPI)~\cite{Sirunyan:2020foa,Bertolini:2014bba} is used to mitigate the effect of pileup at the reconstructed-particle level.
Jet energy corrections are applied to bring the measured response of jets to that of particle level jets on average~\cite{CMS:2016lmd}.
The jet energy resolution amounts typically to 15--20\% at 30\GeV, 10\% at 100\GeV, and 5\% at 1\TeV~\cite{CMS:2016lmd}.
Jets (large-radius jets) overlapping within a cone of radius 0.4 (0.8) in the $\eta$-$\phi$ plane with a muon or an electron passing the criteria described above are discarded.
For the analysis, jets and large-radius jets with $\abs{\eta}<2.4$ as well as with $\pt>25\GeV$ and $\pt>200\GeV$, respectively, are selected.

In addition, jets with $\abs{\eta}>2.4$ (forward jets) are used in the identification of \HH production via the \VBF process, which is characterised by the presence of jets with large $\abs{\eta}$ values.
Jets that are less (more) forward with $2.4<\abs{\eta}<3$ ($\abs{\eta}>3$) and $\pt>50\GeV$ ($\pt>30\GeV$) are selected.
Out of all the jets and the forward jets that are not used for the
\Htobb reconstruction, all possible pairs are selected, and the pair
with the largest pseudorapidity gap $\Delta\eta$ is referred to as the \VBF jet pair.
An event is considered as \VBF tagged if the \VBF jet pair has $\Delta\eta>3$ and an invariant dijet mass larger than 500\GeV.

Jets arising from the hadronisation of \PQb quarks (\PQb jets) are
identified using the \ParticleNet tagging algorithm~\cite{Qu:2019gqs,CMS-DP-2024-024,CMS-DP-2024-066}.
Jets are considered as \PQb tagged if they fulfil a requirement on the
\ParticleNet discriminant that corresponds to a \PQb tagging efficiency of approximately
80\% and a 1\% (14\%) misidentification probability for light-flavour
quark and gluon (charm quark) induced jets.
The \ParticleNet algorithm is further used to identify Higgs bosons with large \pt decaying to \bbbar~\cite{Qu:2019gqs,CMS:2025kje}.
Large-radius jets are considered as boosted \Htobb candidate jets if they fulfil a requirement on the \ParticleNet discriminant that corresponds to an efficiency of approximately 26\% (22\%) of identifying \Htobb jets for \ggF (\VBF) production, with a background misidentification rate of 2\%.
The \PQb jet~\cite{BTV-16-002,CMS-DP-2024-025} and the \Htobb
jet~\cite{CMS:2025kje} (mis)identification probabilities are corrected
to match the efficiencies measured in data.

\section{Data and simulated event samples}\label{sec:samples}
Data from \pp collisions at $\sqrt{s}=13.6\TeV$ collected in 2022 and 2023, corresponding to an integrated luminosity of \lumitot, are analysed.
The data are reconstructed separately in four different data-taking periods with different detector conditions---8\fbinv and 27\fbinv of the data collected in 2022 as well as 18\fbinv and 10\fbinv of the data collected in 2023---and appropriate per-period calibrations are applied before the data are combined for the analysis.
The simulated events described in the following are generated accordingly in four different samples for each physical process, reflecting the different data-taking conditions.

Signal and background events are modelled using the Monte Carlo (MC) event generators \POWHEG~v2.0~\cite{Nason:2004rx,Frixione:2007vw,Alioli:2010xd,Jezo:2015aia} or \MGvATNLO~v2.9.13~\cite{Alwall:2014hca}, interfaced with a detailed detector simulation based on \GEANTfour~\cite{Agostinelli:2002hh}.
The proton structure is described by the parton distribution function (PDF) set {NNPDF3.1}~\cite{Ball:2017nwa} at next-to-next-to-leading order (NNLO) precision in quantum chromodynamics (QCD) perturbation theory.
Parton showering (PS) and hadronisation are simulated with \PYTHIA~v8.306~\cite{Bierlich:2022pfr}, where the parameters for the underlying event description correspond to the CP5 tune~\cite{Sirunyan:2019dfx}.
For comparison with the observed distributions, the events in the simulated samples are normalised to the same integrated luminosity as the data sample, according to their predicted cross sections.
Pileup effects are modelled by adding simulated minimum bias events to all simulated events with a frequency distribution adjusted to match that observed in the data.

The \HH signal process in the \ggF production channel is simulated at next-to-leading order (NLO) precision in QCD with \POWHEG~\cite{Bagnaschi:2011tu,Heinrich:2017kxx}.
Samples with different values of \kappaL are generated, allowing the kinematic distributions for different \kappaL values to be modelled using a weighted sum of the samples~\cite{Heinrich:2019bkc}.
Signal events in the \VBF production channel are generated with \MGvATNLO at leading-order (LO) precision.
Samples with different values of \kappaL and \kappaVV are produced, allowing the kinematic distributions to be modelled for each (\kappaL, \kappaVV) configuration, following the same principle as for the \ggF samples.
The resulting inclusive \HH cross section for each value of the coupling modifiers is normalised to a calculation at NNLO precision in QCD~\cite{Grazzini:2018bsd,Karlberg:2024zxx,Baglio:2020wgt}.
All relevant \HH decay modes are considered, with relevant contributions arising from \bbWW, $\bbbar\PGt\PGt$, and $\bbbar\PZ\PZ$.

The final state targeted by this search is dominated by background processes from top quark-antiquark (\ttbar) and single top quark (\singlet) production, as well as $\PZ/\PGg^{\ast}+\text{jets}$, referred to as Drell--Yan (\DY) production.
The overall normalisation of the \ttbar and \DY processes are measured in situ in the final fit to data described in Section~\ref{sec:results}.

The \ttbar background process is simulated at NLO precision in QCD using \POWHEG~\cite{Frixione:2007nw}.
The top quark \pt spectrum in the \POWHEG simulation has been found to be harder than the one observed in data. This is attributed to missing higher-order calculations~\cite{CMS:2018htd,CMS:2024ybg}, and therefore the simulated events are reweighted to match the prediction at NNLO precision in QCD and NLO precision in electroweak corrections~\cite{Czakon:2017wor}.
The corrections at 13\TeV are taken from Ref.~\cite{Czakon:2017wor}, and an additional small correction is applied to extrapolate them to 13.6\TeV, following Ref.~\cite{CMS:2024okz}.
The inclusive cross section is normalised to a calculation at NNLO precision in QCD including resummation of next-to-next-to-leading logarithmic soft-gluon terms~\cite{bib:ttbarxs:1,bib:ttbarxs:2,bib:ttbarxs:3,bib:ttbarxs:4,bib:ttbarxs:5,bib:ttbarxs:6,bib:ttbarxs:7}.

The \singlet background is dominated by $\PQt\PW$ production, with
small additional contributions from single top quark production in the
$t$ and $s$ channels.
The processes are simulated at NLO precision in QCD, for $\PQt\PW$
production and in the $t$ channel with \POWHEG~\cite{powheg:singlet:st,powheg:singlet:tW}, and in the $s$ channel with \MGvATNLO.
The inclusive cross section of the $\PQt\PW$ process is normalised to
calculations at approximate NNLO precision~\cite{bib:singletxs:tw:1},
and the $t$ and $s$ channel processes to predictions at NLO precision~\cite{bib:singletxs:t:1,bib:singletxs:t:2}.

The \DY background process is simulated with up to two partons at the matrix element (ME) level at NLO precision in QCD with \MGvATNLO, where the merging of the jets from the ME calculation and the PS is done using the {\scshape FxFx}~\cite{Frederix:2012ps} prescription.
Dedicated corrections, described later in Section~\ref{sec:strategy:bkgmodelling}, are derived from data to improve the modelling of relevant observables, such as the jet multiplicity spectrum in \DY events.
The \DY events are separated into two mutually exclusive processes, based on the flavour of the jets at the particle level, where these jets fulfil the acceptance requirements of $\pt>20\GeV$ and of $\abs{\eta}<2.4$.
Events with at least one particle-level jet that contains one or more \PQb- or \PQc-flavoured hadrons are classified as $\DY+\text{heavy-flavour}$ jets (\DYHF) events, and all other events are classified as $\DY+\text{light-flavour}$ and gluon jets (\DYLF) events.

Minor background contributions arise from the resonant production of
single Higgs bosons (\PH production).
The \PH production processes in the \ggF and \VBF channels, associated production with a
vector boson ($\PV\PH$), $\PZ\PH$ production in \ggF, and \ttbar
associated production are simulated at NLO precision using
\POWHEG~\cite{Alioli:2008tz,Nason:2009ai,Mimasu:2015nqa,Jones:2018hbb,Hartanto:2015uka}.
Associated $\ttbar\PV\PH$ production is simulated at NLO precision
using \MGvATNLO, and \singlet associated \PH production
in the $\PQt\PH\PQq$ and $\PQt\PH\PW$ channels is simulated at LO
precision using \MGvATNLO.
Their cross sections are normalised to the latest available
calculations, taken from Ref.~\cite{Karlberg:2024zxx}.

Further minor background contributions stem from \Wjets production, the production of two (\diboson) or three (\triboson) \PZ or \PW bosons, \ttbar production in association with one (\ttV) or two (\ttVV) \PZ or \PW bosons, and from the production of four top quarks (\tttt).
They are simulated using \POWHEG~\cite{Nason:2013ydw} (\diboson, \triboson) and \MGvATNLO
(\Wjets, \ttV, \ttVV, \tttt).

\section{Event selection and analysis strategy}\label{sec:strategy}
Events are selected with exactly two oppositely charged leptons (\mumu, \ee, \emu).
The selected events are categorised into several signal regions (SRs),
targeting \ggF and \VBF production, or into background
control regions (CRs), based on
the output of a multiclassification NN as well as the
\PQb-tagged jets multiplicity and the presence of a boosted \Htobb candidate jet.
The events in the SRs are classified by additional NNs designed to separate \ggF or
\VBF signal events from background events.
The distributions of the output values of the binary NNs in the SRs, together with the
event yield in the CRs, enter the final fit as sensitive observables.

Contributions from SM background processes are modelled using
the simulated events described in Section~\ref{sec:samples}.
Dedicated corrections are derived for the \DY events to improve the
modelling of the data, and the normalisations of the dominant \ttbar
and \DY background processes are determined in the final fit to data.

\begin{figure}[!hpt]
  \centering
  \includegraphics[angle=90, width=0.64\textwidth]{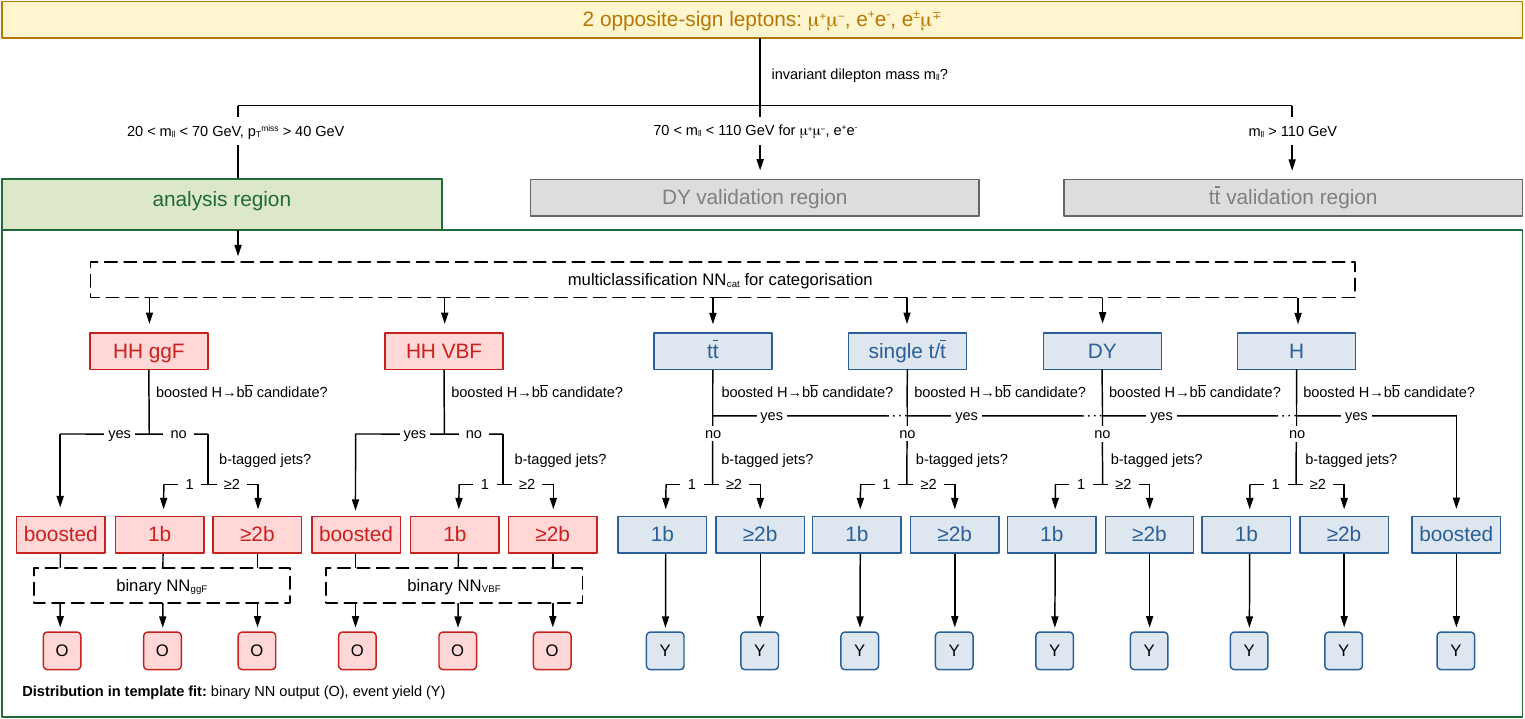}
  \caption{
    Illustration of the event categorisation:
    SRs are depicted in red, background CRs in blue.
    Details of the NNs are described in the text.
    The binary NN output distributions (O) and the event yields (Y) in the CRs enter the final fit as sensitive observables.
  }
  \label{fig:analysis_strategy}
\end{figure}

The event selection, the analysis strategy, the details of the NN
architecture and training, as well as the background
modelling are detailed in the following, and the analysis strategy is
illustrated in Fig.~\ref{fig:analysis_strategy}.

\subsection{Event selection}
The data are selected using a combination of triggers.
The triggers require the presence of either one or two muon candidates
for the \mumu events, one or two electron candidates for the \ee
events, as well as one muon or one electron candidate, or one muon and
one electron candidate, for the \emu events.
The minimal thresholds on the trigger objects are $\pt>8\GeV$ for the
muons and $\pt>12\GeV$ for the electrons.
The trigger selection for \ee and \emu events features an
inefficiency at lepton \pt above 100\GeV owing to an electron
isolation criteria employed in the electron triggers.
This is compensated by using additional triggers with looser electron
isolation criteria that require the presence of two electrons with
$\pt>33\GeV$ or of one electron with $\pt>50\GeV$ and one jet with
$\pt>165\GeV$.
The latter criteria are new compared to the previous search~\cite{CMS:2024rgy}.
As a result, the total selection efficiency with respect to the offline reconstruction achieved for signal events after the event
selection described below is 98\% in the \mumu channel, 95\% in the \ee
channel, and 96\% in the \emu channel.
The single-lepton triggers and the two new triggers contribute up to 12\% to the signal selection efficiency, depending on the channel.

A further offline selection, referred to as the baseline selection, requires events to have exactly two
muons or two electrons or one muon and one electron of opposite
charge, and at least one \PQb-tagged jet.
Events are rejected if further muons or electrons are present with
$\pt>5$ or 7\GeV, respectively.
Events must be accepted by a trigger based on online leptons with a
flavour consistent with that of the two leptons selected
offline; for example, \mumu events are required to pass a dimuon or
single-muon trigger.
The invariant mass of the selected lepton pair, \mll, is required to
be larger than 20\GeV to suppress events from heavy-flavour resonance
decays and low-mass \DY processes.

The selected events are then separated into an analysis region with
$20<\mll<70\GeV$ and $\ptmiss>40\GeV$ as well as into validation
regions with $\mll>110\GeV$, enriched in \ttbar events, and with $70 <
\mll < 110\GeV$ for \ee and \mumu events, enriched in \DY events.
The analysis region events are used in the final analysis to extract
the signal, and the validation region events are used to verify the
\ttbar and \DY background modelling and derive corrections for the \DY
background.
The total selection efficiency for $\HH\to\bbWW$ signal events in final states with two leptons (\ee, \mumu, or \emu) in the analysis region is 12\%.

The event selection criteria are summarised in
Table~\ref{tab:baselineselection}.
\begin{table}[!htbp]
  \centering
  \topcaption{
    Event selection criteria in the analysis region (AR) and the \DY
    and \ttbar validation regions (VR).
  }
  \label{tab:baselineselection}
  \renewcommand{\arraystretch}{1.3}
  \begin{tabular}{lccc}
    & \multicolumn{3}{c}{Baseline selection} \\
    \hline
    Number of leptons              & \multicolumn{3}{c}{2} \\
    \pt of leading lepton          & \multicolumn{3}{c}{$>$25\GeV} \\
    \pt of subleading lepton       & \multicolumn{3}{c}{$>$15\GeV} \\
    $\abs{\eta}$ of electrons      & \multicolumn{3}{c}{$<$2.5} \\
    $\abs{\eta}$ of muons          & \multicolumn{3}{c}{$<$2.4} \\
    Number of jets                 & \multicolumn{3}{c}{$\geq$1} \\
    Number of \PQb-tagged jets     & \multicolumn{3}{c}{$\geq$1} \\
    \pt of jets                    & \multicolumn{3}{c}{$>$25\GeV} \\
    $\abs{\eta}$ of jets           & \multicolumn{3}{c}{$<$2.4} \\[\cmsTabSkip]
    & AR & \DY VR & \ttbar VR \\
    \hline
    Sign and flavour of leptons    & \mumu, \ee, \emu & \mumu, \ee & \mumu, \ee, \emu \\
    \mll                           & 20--70\GeV & 70--110\GeV & $>$110\GeV \\
    \ptmiss                        & $>$40\GeV && \\
  \end{tabular}
  \renewcommand{\arraystretch}{1.0}
\end{table}

\subsection{Analysis strategy}
The events in the analysis region are categorised in several steps, as
illustrated in Fig.~\ref{fig:analysis_strategy}.
First, the events are categorised by a multiclassification NN
(\NNmult) designed to separate between different signal and background
processes.
The values obtained in the output nodes of the \NNmult are normalised to
unity using a ``soft-max'' function, and, as a result, the output
values can be interpreted as probabilities describing the likelihood
of the event being of a certain process.
Events are assigned to the category corresponding to the most probable
process according to this NN multiclassification.
The events are separated into one of two SRs targeting \ggF \HH
production (\SRggF) or \VBF \HH production (\SRVBF), or into
one of four background CRs targeting the main background
processes: \ttbar, \singlet, \DY, and \PH production.

Events in \SRggF and \SRVBF are further sorted into one of three
categories, according to their jet content:
events containing a boosted \Htobb candidate jet are sorted into the ``\catboosted''
categories \SRggFboosted and \SRVBFboosted;
events containing exactly one \PQb-tagged jet are sorted into the ``\catb''
categories \SRggFb and \SRVBFb; and
the remaining events containing two or more \PQb-tagged jets are sorted into the ``\catbb''
categories \SRggFbb and \SRVBFbb.
Events in the CRs are sorted into a dedicated \catboosted category if a boosted
\Htobb candidate jet is present. There is one single such category for all CR
events owing to the low number of events. The remaining CR events are sorted into
the \catb (\catbb) categories if they contain exactly 1 ($\geq$2) \PQb-tagged jets.
This results in 15 categories in total:
three categories (\catboosted, \catb, \catbb) in each of the two SRs (\SRggF,
\SRVBF),
two categories (\catb, \catbb) in each of the four CRs, and a
\catboosted CR.

The events in the three \SRggF categories are further classified by a
dedicated binary NN (\NNggF), which is optimised to separate \ggF
production \HH events from the inclusive background.
The output values of the \NNggF are used as discriminating observables in the
\SRggF categories in the final fit.
Likewise, the events in the three \SRVBF categories are classified by
another binary NN (\NNVBF) optimised to separate \VBF production \HH
events from the background, and the output values of the \NNVBF are
used as final discriminating observables.
The binning in the \SRggF (\SRVBF) categories is chosen such
that the \NNggF (\NNVBF) output distributions are uniform for \ggF
(\VBF) signal events, and such that the uncertainty due to the
limited number of simulated events is sufficiently small.
In the background CRs, the yield in each category enters the final
fit.
The resulting discriminant distribution is shown in Fig.~\ref{fig:classifiers} in Section~\ref{sec:results}.

The analysis strategy has been optimised using simulated events to achieve the best sensitivity.
For example, the inclusion of the dedicated \catboosted categories improves the expected limit on the SM \HH production cross section by 4\% and reduces the confidence interval on \kappaVV by 26\%.

\subsection{Neural network architecture and training}\label{sec:strategy:nn}
The hyperparameters~\cite{ml:general} of the multiclassification \NNmult and the binary \NNggF and \NNVBF are listed in Table~\ref{tab:nn_hyperparameters}.
They have been optimised based on the achieved expected limit to SM
\HH production, evaluated on simulated events.
\begin{table}[hpt]
  \centering
  \topcaption{%
    Hyperparameters of the neural networks.
    Where they differ for the multiclassification \NNmult and the
    binary \NNggF and \NNVBF, they are listed as
    ``\NNmult/\NNggF/\NNVBF'', otherwise they are the same for all
    networks.
  }
  \label{tab:nn_hyperparameters}
  \renewcommand{\arraystretch}{1.3}
  \begin{tabular}{lc}
    \hline
    Hidden layers                & 3 \\
    Nodes per hidden layer       & 512 \\
    Activation function          & ReLU \\
    Output activation function   & Softmax \\
    Loss function                & Categorical cross-entropy \\
    Preprocessing                & Batch normalisation \\
    Dropout rate                 & 20\% \\
    Training epochs              & 100 \\
    Batch size                   & 4032/10648/15972 \\
    Steps per epoch              & 52/23/15 \\
    Initial learning rate        & $5\times10^{-4}$ \\
    Learning rate decay          & 0.8 \\
    Learning rate decay patience & 3 epochs \\
  \end{tabular}
  \renewcommand{\arraystretch}{1.0}
\end{table}

The NNs are trained using simulated signal and background events from
all four analysed data-taking periods together, thereby reducing statistical
fluctuations in the trained NN parameter values.
The events are weighted to reflect the different integrated
luminosities.
The sensitivity of the analysis does not
degrade compared to the case of training NNs separately for each
data-taking period or when using information about the period of data
taking as an input feature to the NNs.

Simulated signal events with different values of \kappaL and \kappaVV
are used for the training.
For \ggF production, four different samples are used with values of
\kappaL of 0, 1, 2.45, and 5, and for \VBF production, six different
samples are used with values of \kappaL and \kappaVV between $-3.39$
and 1 and between 0.959 and 3.57, respectively.
The values have been chosen to represent regions in parameter space
with distinct kinematic properties.
For example, at $\kappaL=0$, there is no contribution from diagrams
involving the trilinear Higgs boson self-coupling
(Fig.~\ref{fig:intro:diagrams} top left) and \HH production proceeds
via the diagram involving only couplings between the top quark and the
Higgs boson (Fig.~\ref{fig:intro:diagrams} top right), while at
$\kappaL=2.45$, the destructive interference between the two types of
diagrams is maximal, such that the \HH production cross section is
minimal.
For the training of the binary NNs, only \HH signal events of the
target production process are considered.
For example, for the training of the \NNggF, only the \ggF production
\HH events are used as signal, while the \VBF production \HH events do
not enter the training.

Events from the different simulated samples are effectively weighted during training to prevent the NN
classification from being biased by the relative frequencies of the underlying processes.
For signal events, the weighting is chosen such that the
different simulated \kappaL and \kappaVV points have the same effective number
of events, obtained as the sum of all weights, in each batch used for training.
For background events in the \NNmult training, the weighting is
chosen such that each background process has the same effective number of
events as the total effective number of signal events.
In the binary NN trainings, the background events
are weighted relative to each other by the cross section of the
processes.
However, a larger relative weight is assigned to the events of some of
the minor background processes in order to avoid large statistical
fluctuations from few events in the analysis bins with largest signal
purity.
Finally, the total effective number of background events is normalised
to the total effective number of signal events.
This scheme has been chosen in order to find a compromise between
sensitivity to the SM \HH signal, to \kappaL, and to \kappaVV,
evaluated by the expected limits, as well as to minimise the size of
MC statistical uncertainties.

A five-fold cross-validation technique is used for the training and
evaluation of the NNs.
In each fold, the 80\% of events used for training are further split into three independent
subsamples used for the actual training (75\%), for the
optimisation of the hyperparameters (10\%), and for validating the
performance of the NNs (15\%).
The training is terminated after 100 epochs.

\begin{figure}[!hpt]
  \centering
  \begin{tabular}{@{}c@{}c@{}}
    \includegraphics[width=0.45\textwidth]{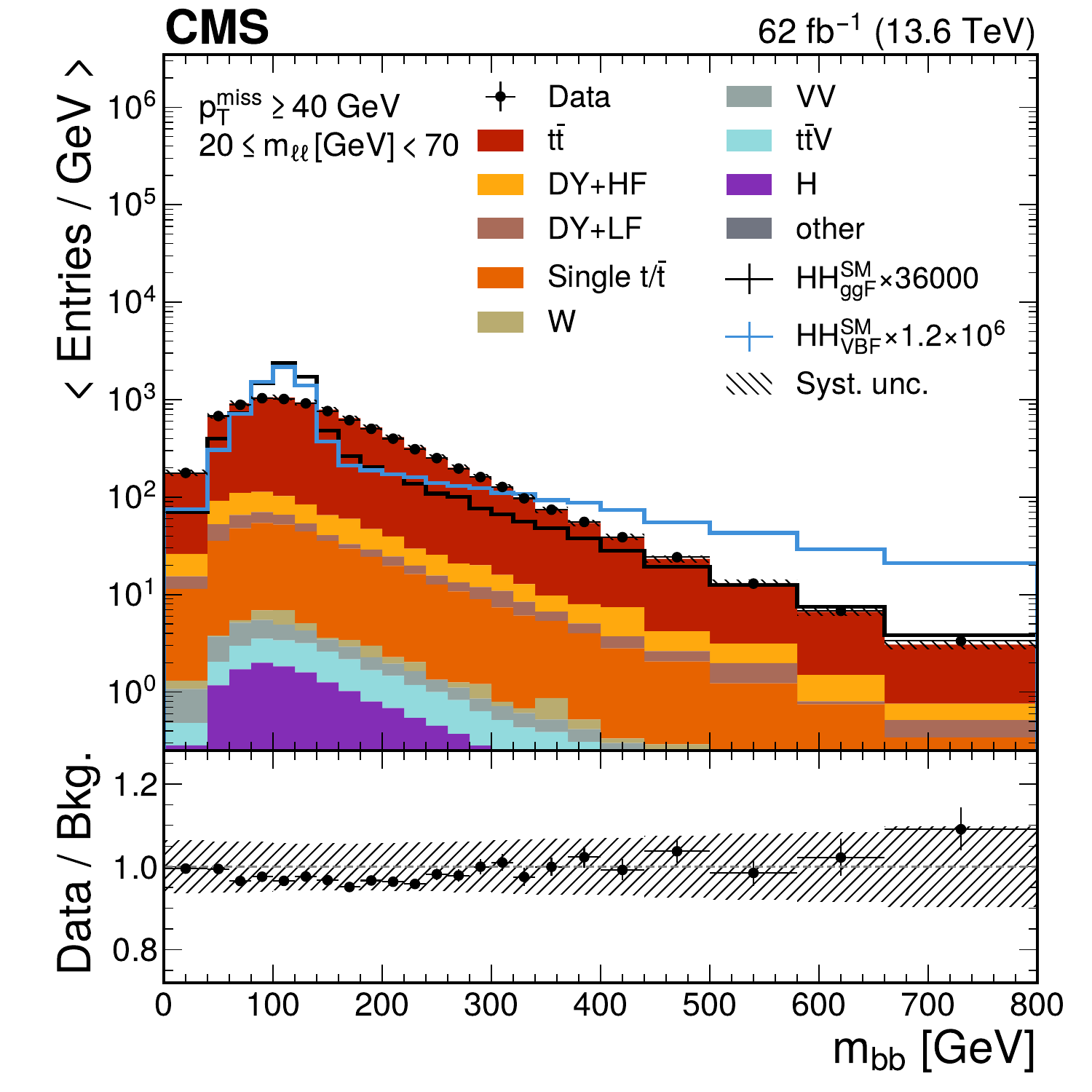} &
    \includegraphics[width=0.45\textwidth]{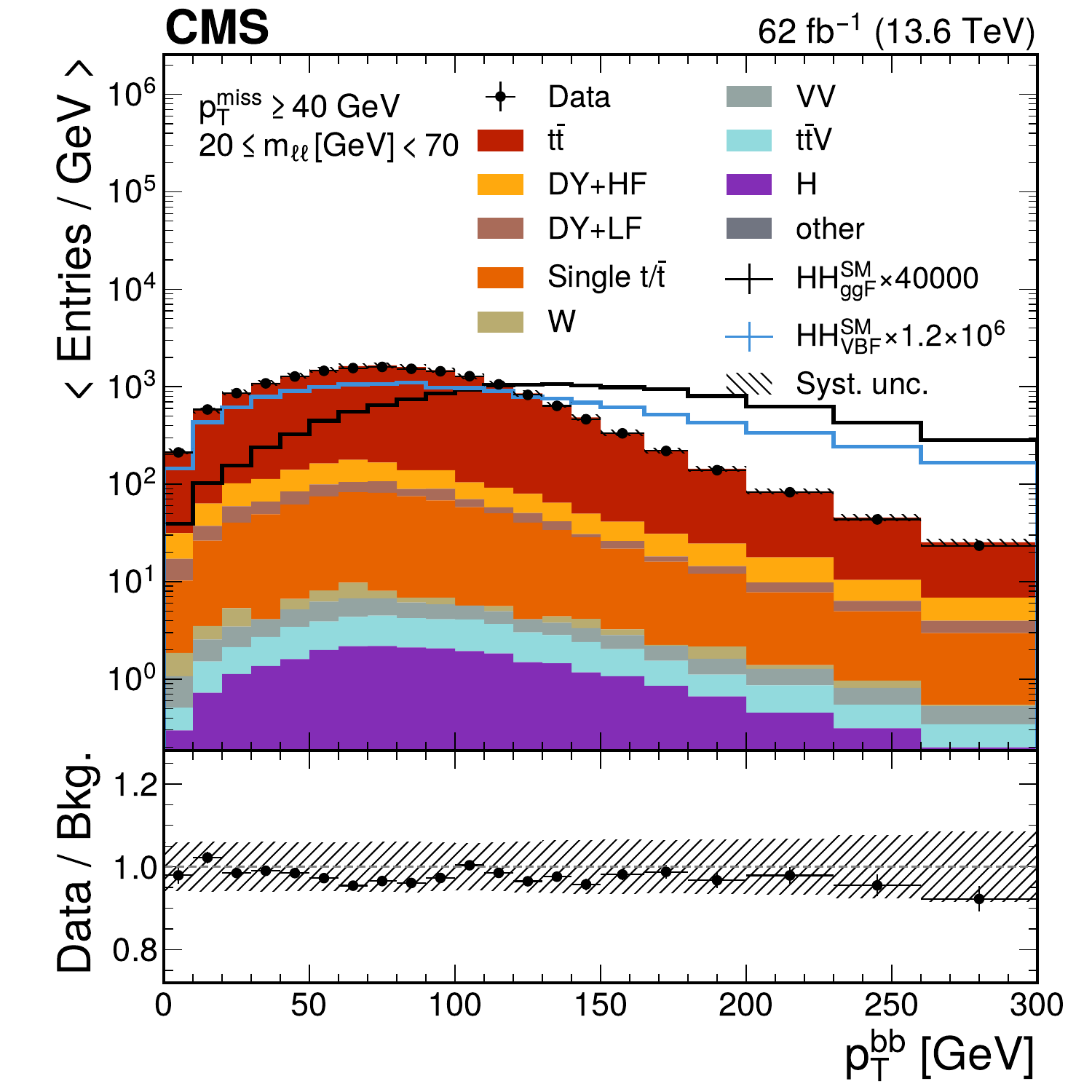} \\
    \includegraphics[width=0.45\textwidth]{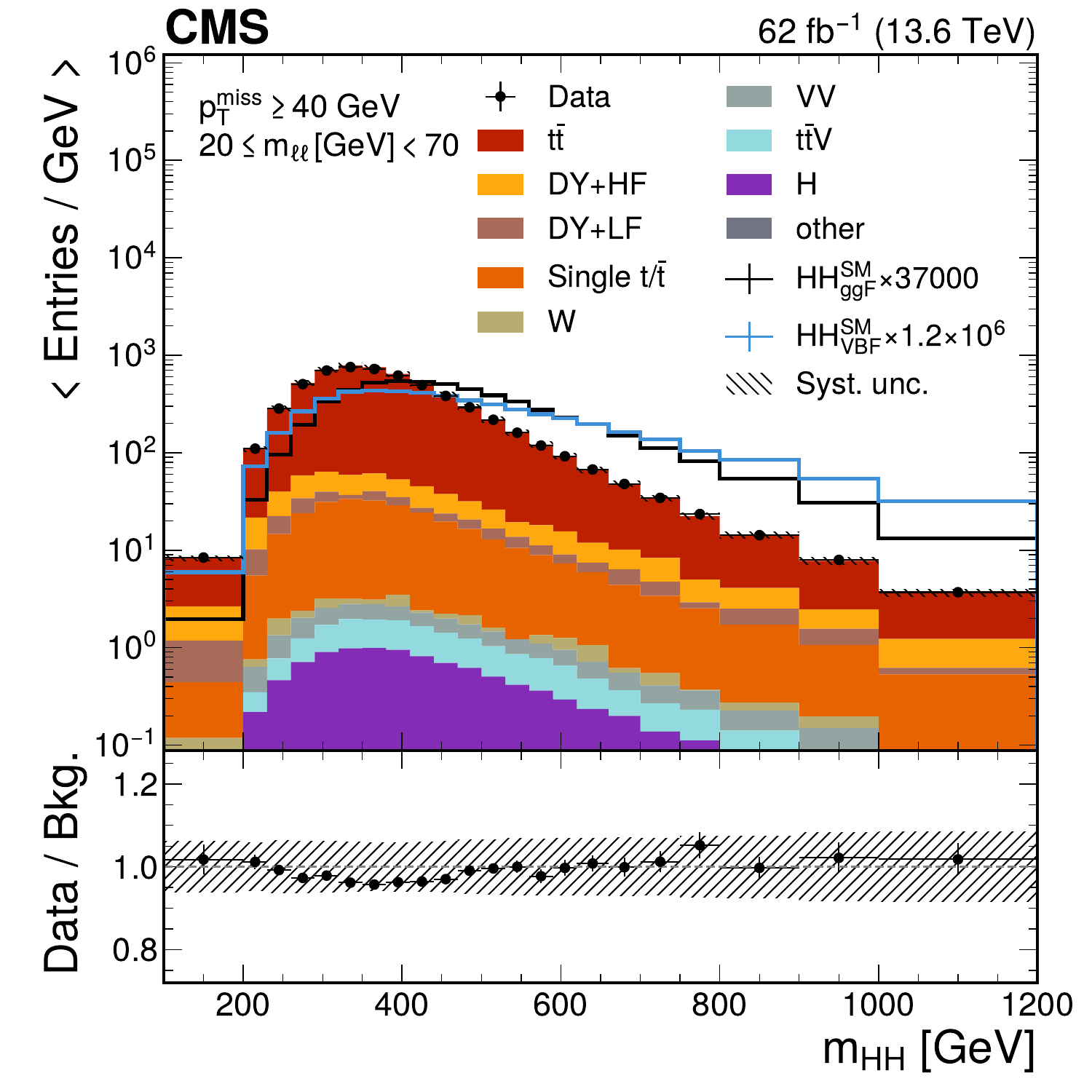} &
    \includegraphics[width=0.45\textwidth]{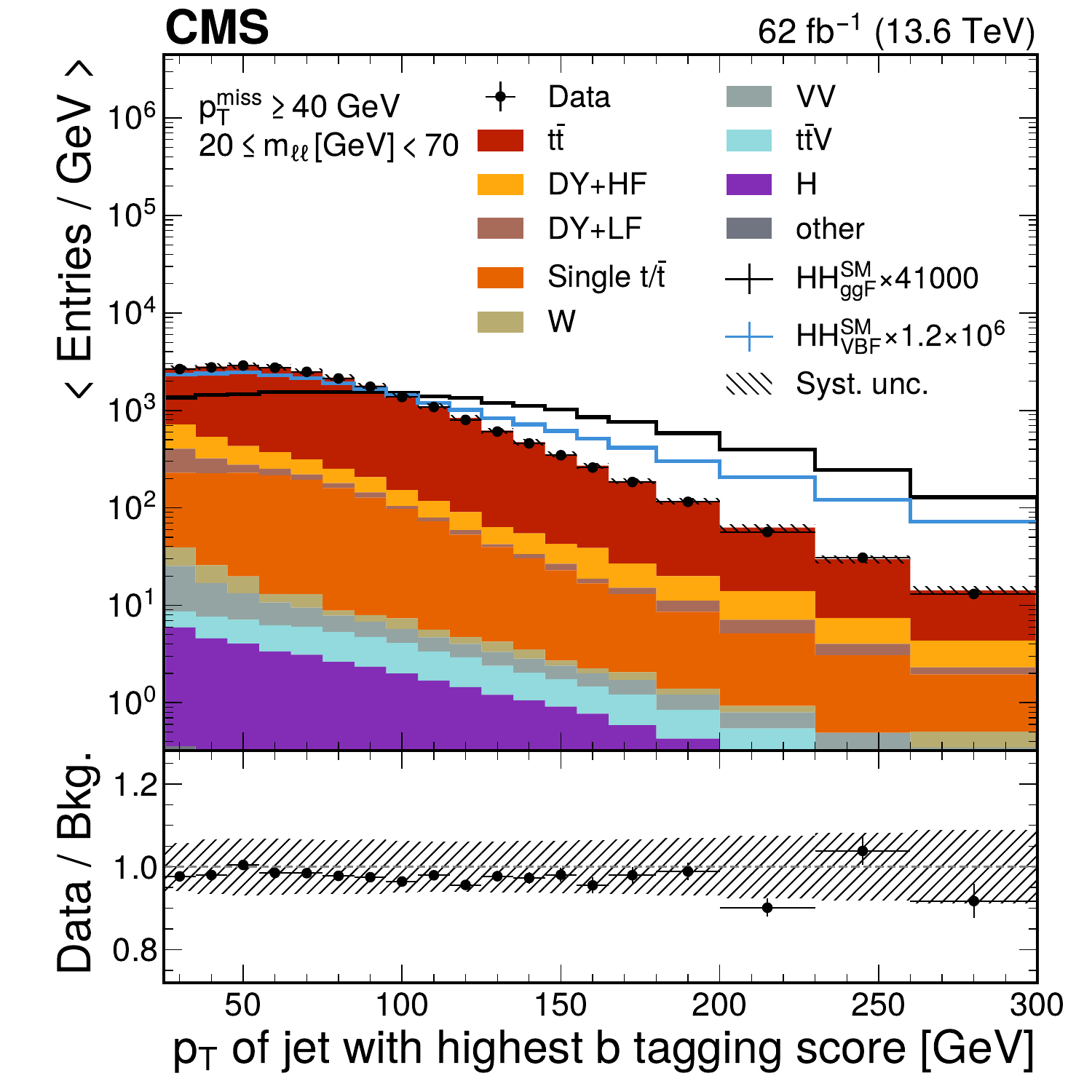} \\
  \end{tabular}
  \caption{%
    Invariant mass (upper left) and \pt (upper right) of the
    \PH candidate decaying to \bbbar, reconstructed as the invariant mass and \pt, respectively, of the two jets with the highest \PQb tagging score;
    invariant mass of the \HH system (lower left), reconstructed as the invariant mass of the two jets with the highest \PQb tagging score, the two leptons, and \ptmiss;
    and \pt of the jet with the highest \PQb tagging score (lower right),
    for events in the analysis region observed in data (markers) and predicted by the
    background model (stacked histograms) prior to the fit to data.
    The \HH signal distributions in the \ggF and \VBF
    production channels as predicted in the SM, scaled to the total background yield for
    better visibility, are overlaid (solid lines).
    The uncertainty band represents the total systematic uncertainty.
  }
  \label{fig:nn_inputvariables}
\end{figure}

The input variables of the NNs are listed in
Table~\ref{tab:nn_inputvariables}.
The same variables are used in the multiclassification and in the
binary NNs, with the exception of the \VBF tag information, which only
enters the \NNVBF.
No dedicated input variables based on the properties of the boosted \Htobb candidate jets are considered.
The input variables have been chosen out of a larger set of possible
observables as the ones that are most important for the performance of
the NNs, based on several ranking procedures including a ``data
Shapley'' metric~\cite{ghorbani2019data} and the expected limit on the
SM \HH production cross section.
The exact importance ranking of each variable varies for the different NNs.
As an example, four input variables with generally high importance based on the data
Shapley metric are shown in Fig.~\ref{fig:nn_inputvariables}:
the invariant mass and the \pt of the \PH candidate decaying to \bbbar, reconstructed as the invariant mass and \pt, respectively, of the two jets with the highest \PQb tagging score;
the invariant mass of the \HH system, reconstructed as the invariant mass of the two jets with the highest \PQb tagging score, the two leptons, and \ptmiss;
and the \pt of the jet with the highest \PQb tagging score.
\begin{table}[!!!t]
  \centering
  \topcaption{
    Observables used as input variables to the NNs.
  }
  \renewcommand{\arraystretch}{1.4}
  \begin{tabular}{p{0.9\textwidth}}
    \hline
    Invariant mass of \bbbar system$^{\dagger}$; proxy for
    the mass of the \Htobb candidate\\
    Invariant mass of $\Pell\Pell$ system (\mll)\\
    Invariant mass of the leptons and \ptmiss; proxy for the mass of
    the $\PH\to\PW\PW$ candidate\\
    Invariant mass of the \PQb jets$^{\dagger}$, leptons, and
    \ptmiss; proxy for the mass of the \HH system\\[\cmsTabSkip]
    Number of jets\\
    Scalar sum of jet \pt (\HT)\\
    \pt of jet with highest \PQb tagging score\\
    \pt of jet with second-highest \PQb tagging score\\
    \pt of leading jet that is not considered as part of the \bbbar system$^{\dagger}$ \\
    \pt of the \bbbar system$^{\dagger}$\\
    Sum of \PQb tagging scores of all jets\\[\cmsTabSkip]
    \pt of leading lepton\\
    \pt of subleading lepton\\
    \pt of the $\Pell\Pell$ system\\
    $\eta$ of leading jet that is not considered as part of the \bbbar system$^{\dagger}$ \\[\cmsTabSkip]
    Maximum $\Delta R$ between jets\\
    $\Delta R$ between jets of \bbbar system$^{\dagger}$\\
    $\Delta R$ between leptons\\
    $\Delta R$ between \bbbar system$^{\dagger}$ and $\Pell\Pell$ system\\
    Minimum $\Delta R$ between lepton and \PQb-tagged jet\\[\cmsTabSkip]
    \ptmiss\\
    Binary tag of lepton flavour combinations: same flavour or mixed flavour\\
    Binary \VBF tag based on the presence of a \VBF jet pair$\;^{\ddagger}$\\
    \\
    $^{\dagger}$ two jets with highest \PQb tagging scores; in
    events with only one jet, the observable is assigned a
    non-physical default value of $-10$\\[\cmsTabSkip]
    $^{\ddagger}$ only used in \NNVBF
  \end{tabular}
  \renewcommand{\arraystretch}{1.0}
  \label{tab:nn_inputvariables}
\end{table}

\subsection{Background modelling}\label{sec:strategy:bkgmodelling}
All background processes are modelled using the simulated event
samples described in Section~\ref{sec:samples}, with the corrections
and calibrations discussed in the previous sections.
The overall normalisation of the \ttbar and the \DY background
processes are free parameters in the final fit to data.
Two separate \ttbar normalisation parameters are assigned, one for
the \catb and \catbb categories, and one for the \catboosted categories,
in order to account for potential modelling differences in the two very
different \pt regimes.
For the \DY background, two separate normalisation parameters are
assigned, one to the \DYHF and one to the \DYLF process, to account for
flavour-dependent effects in the hadronic part of the events.

The background modelling of the dominant \ttbar and \DY background
processes is verified in the \ttbar validation region with $\mll>110\GeV$
and in the \DY validation region with $70 < \mll < 110\GeV$ for \ee
and \mumu events, respectively.
While the relevant observables are sufficiently well modelled for \ttbar events, a mismodelling is observed for \DY events, in particular in the distribution of the jet multiplicity, \Njets , as well as in the \pt spectrum of the dilepton system, \ptll.
Therefore, a dedicated correction of the \DY events is derived as a function of \ptll and \Njets, using the data in the \DY validation region.

\begin{figure}[!hpt]
  \centering
  \begin{tabular}{@{}c@{}c@{}}
  \includegraphics[width=0.45\textwidth]{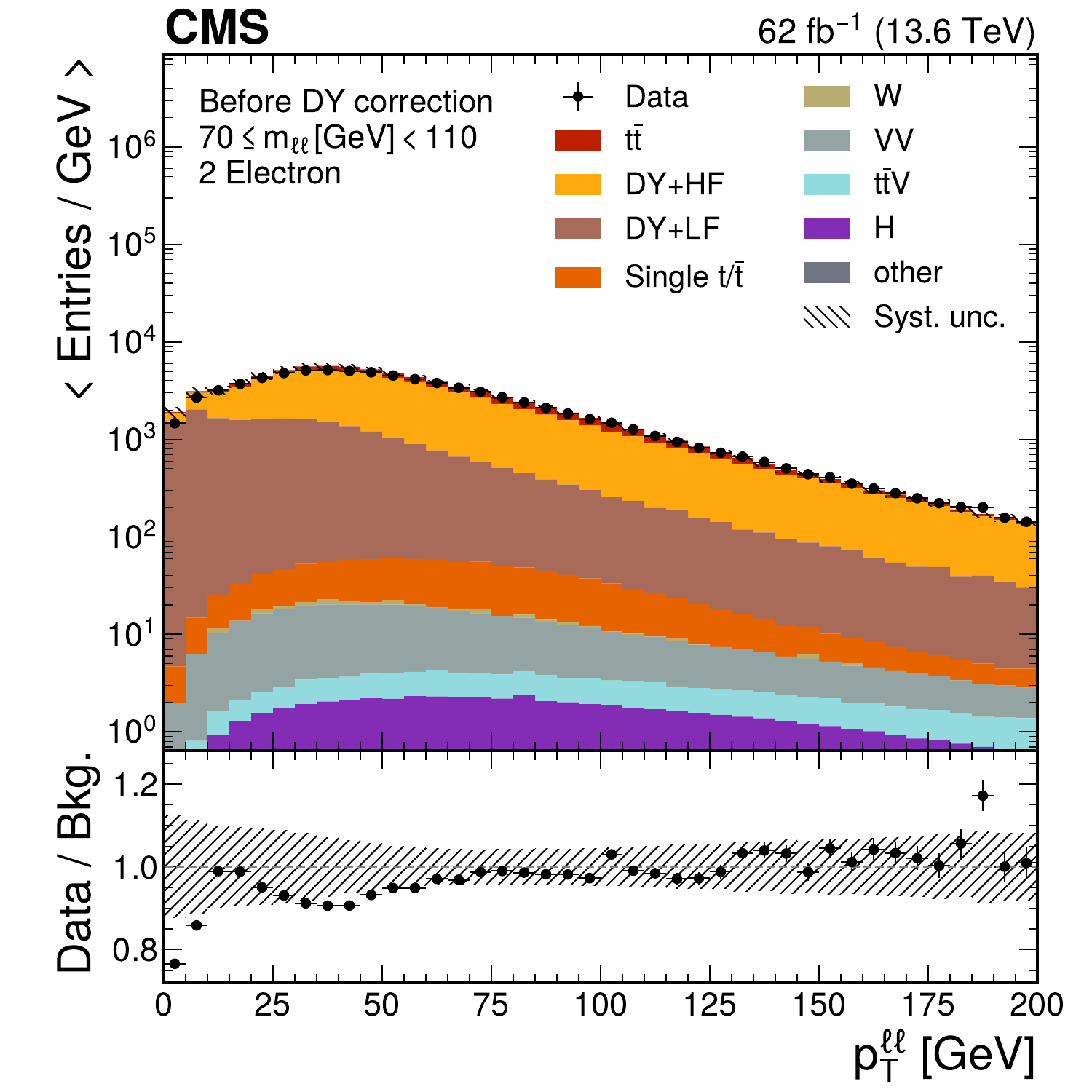} &
  \includegraphics[width=0.45\textwidth]{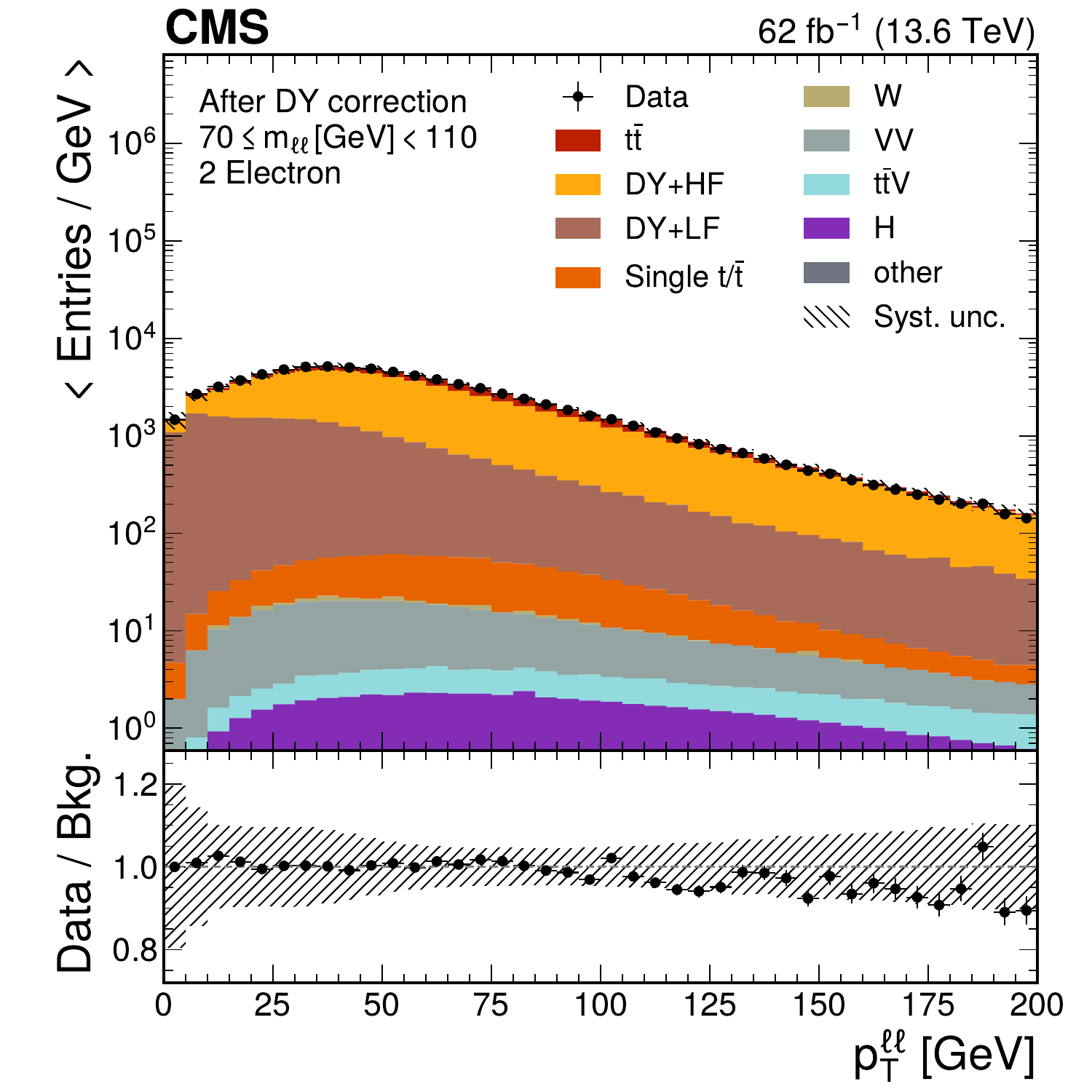}
    \\
  \end{tabular}
  \caption{%
    The \pt of the dilepton system in the \ee channel in the DY validation region before
    (left) and after (right) application of the DY corrections.
    The uncertainty band shows the total systematic uncertainty.
  }
  \label{fig:dycorrection}
\end{figure}

The \DY correction is determined from \mumu events in the \DY validation region, which has a purity of 94\%, using the following procedure.
The \ptmumu distribution observed in data, after subtraction of the expected non-\DY contribution predicted by simulation, is divided by the \ptmumu distribution predicted by the \DY simulation, and the resulting ratio is parameterised using an empirical function $\fDYcorr(\ptll)$.
Then, \fDYcorr is applied as an event weight to the simulated \DY events in the analysis region as a function of the particle-level \ptll, correcting the simulated \ptll spectrum to match the data.
Applying the correction as a function of the particle-level \ptll avoids dilution due to the electron momentum resolution, while for the muons the effect of the \pt resolution is negligible.
The correction \fDYcorr is derived separately for events with $\Njets=1$ and $\Njets\geq2$, and for the latter, an additional normalisation factor is applied per \Njets bin to account for residual mismodelling of the jet multiplicity.
The correction derived in the \DY validation region is applicable also in the analysis region, since the relative fraction of \DYHF and \DYLF  events is the same in both regions.
The effect of the corrections is demonstrated in
Fig.~\ref{fig:dycorrection}, which shows the \ptee distribution in \ee
events in the DY validation region before and after application of the
corrections.
The robustness of the correction procedure has been carefully
verified, for example by testing different correction functions and by
deriving dedicated corrections per data-taking period, leading to
results consistent with those of the nominal method.

\section{Systematic uncertainties}\label{sec:systematics}
The sensitivity of the analysis is primarily limited by statistical uncertainties.
However, several sources of systematic uncertainties also affect the result and have been considered as described in the following.
The systematic uncertainties alter either the rate or both the rate
and the shape of the discriminant distributions for the
signal or background processes.
Their impact on the measured signal and background yields is discussed in Section~\ref{sec:results}.

Theoretical uncertainties are treated as fully correlated among the
different data-taking periods, while experimental uncertainties are in
most cases treated as uncorrelated between the periods.
The latter assumption is motivated by the fact that the experimental uncertainties are
mainly of statistical origin related to the limited size of the data
and simulation samples used in auxiliary measurements, which are
independent across the data-taking periods, and that the detector
conditions changed between the data-taking periods.

Uncertainties in the inclusive cross sections used to
predict the rate of the signal and background processes arise primarily
from variations of the renormalisation and factorisation scales and
the PDFs, as well as the treatment of the top quark mass in case of the
signal in the \ggF production channel, and are propagated to the yield estimates.
The cross section uncertainties are each separated into a component
($\sigma(\muR/\muF)$) due to the renormalisation (\muR) and
factorisation (\muF) scales, or scales and top quark mass
($\sigma(\muR/\muF+m_{\PQt})$) for signal, and a component due to the PDF ($\sigma(\text{PDF}+\alpS)$).
The uncertainties are correlated where appropriate among processes;
for example, for processes with the same dominant initial state, such
as $\ttbar\PZ$ and \tttt, the PDF uncertainties are treated as correlated.
Since the normalisation of the \ttbar and \DY background
processes are measured in situ, they are not
affected by the uncertainties of the inclusive cross section
predictions.
For the signal, the uncertainties in the inclusive cross section
prediction generally affect the result since the measured rate is
normalised to the SM expectation.

Uncertainties in the simulation of the signal and background events
are evaluated as follows.
Uncertainties arising from missing higher-order terms at the ME level
are evaluated from variations of the \muR and \muF
scales by factors of 2 and 0.5 with respect to the nominal values.
All possible permutations of independent scale variations are
performed, excluding the two extreme variations where one scale is
varied by a factor of 2 and the other by a factor of 0.5, and the envelope of the six obtained variations in the fitted distributions is assigned as the uncertainty (\muRF scale)~\cite{Cacciari:2003fi}.
The uncertainty originating from the PDF set is determined from the PDF
variations provided with the {NNPDF3.1} set~\cite{Ball:2017nwa},
following the procedure described in Ref.~\cite{PDF4LHCWorkingGroup:2022cjn},
correlating processes for which the same flavour scheme and order in \alpS are
used in the PDF set.
The uncertainty in the \PYTHIA PS is determined by varying the
parameters controlling the amount of initial- and final-state
radiation independently by factors of 2 and 0.5 with respect to their
nominal value~\cite{Bierlich:2022pfr}.
The initial-state radiation uncertainties are treated as correlated
among the processes, while the final-state radiation uncertainties are
treated as uncorrelated between the processes.
Uncertainties in the corrections applied to the \pt spectrum of the
top quarks in the \ttbar simulation~\cite{CMS:2018htd} are also
considered; they amount up to 7\% of the expected yield depending on the \pt of the top
quarks.

The theory modelling uncertainties described in the previous paragraph
arising in the simulation of the events can affect the overall
normalisation, the detector acceptance, and the shapes of the fitted
distributions.
For the \ttbar and \DY background processes, whose
normalisations are determined in the fit, the impact of these
variations on the inclusive normalisation (\ie, before any event
selection) is removed, leaving only acceptance and shape
effects.
Likewise, for the other processes, the variation in the
normalisation is removed in order to avoid double counting with the
uncertainty in the inclusive cross section prediction used to predict
the rates of the processes.

Effects of the uncertainty in the distribution of the number of pileup
interactions are evaluated by varying the total inelastic cross
section used to predict the number of pileup interactions in the
simulated events by $\pm4.6\%$ from its nominal value~\cite{Sirunyan:2020foa}.
The uncertainties in the integrated luminosity used to normalise the
predicted signal and background yields are 1.4\% for
2022~\cite{CMS-PAS-LUM-22-001} and 1.3\% for
2023~\cite{CMS-DP-2024-068}.

The trigger efficiency is measured in an unbiased sample of events,
mostly \ttbar leptonic decays, that contain two leptons and fulfil
the offline event selection criteria but were collected with triggers
based on \ptmiss, which feature a negligible correlation with the
triggers used in this analysis.
Corrections are derived to match the efficiency in simulation to that
observed in data, with uncertainties of up to 5\%, dominated by
statistical fluctuations in the data samples used in the measurement.

Residual differences between the lepton reconstruction,
identification, and isolation efficiencies in data and simulation are
corrected based on efficiency measurements using a ``tag-and-probe''
method in data samples of $\PZ\to\Pell\Pell$
events~\cite{CMS:2020uim,CMS:2018rym}.
The uncertainties are smaller than 1 and 2\% for the muon and electron
correction factors, respectively.

The uncertainties in the jet energy scale and resolution are propagated
by varying the corrections applied to all jets in the
signal and background simulations within their uncertainty~\cite{CMS:2016lmd}.
The uncertainties in the corrections of the \PQb jet (mis)identification probabilities in simulated events~\cite{BTV-16-002,CMS-DP-2024-025}, which can arise from
the contamination of background processes in the data samples used for
their derivation, are considered separately for light-flavour quark
and gluon jets, \PQb jets, and \PQc jets, and vary between 5 and 30\%, depending on the jet \pt.
Uncertainties in the correction of the boosted \Htobb candidate jet tagging
efficiency amount to approximately 5\%, and are considered separately
for the \catboosted \ggF and \VBF SRs and the \catboosted CR.

The uncertainty in the \DY spectrum correction is evaluated as half
of the difference between the nominal and the corrected distributions,
and applied symmetrically to the corrected distributions.
It has been verified that the size of the uncertainty does not bias the result of the analysis.

The impact of statistical fluctuations in the signal and background
prediction due to the limited number of simulated events (MC stat unc.) is accounted
for using the Barlow--Beeston ``lite'' approach~\cite{Conway:2011in}.

\section{Results}\label{sec:results}
\begin{figure}[htp]%
  \centering%
  \includegraphics[width=0.9\textwidth]{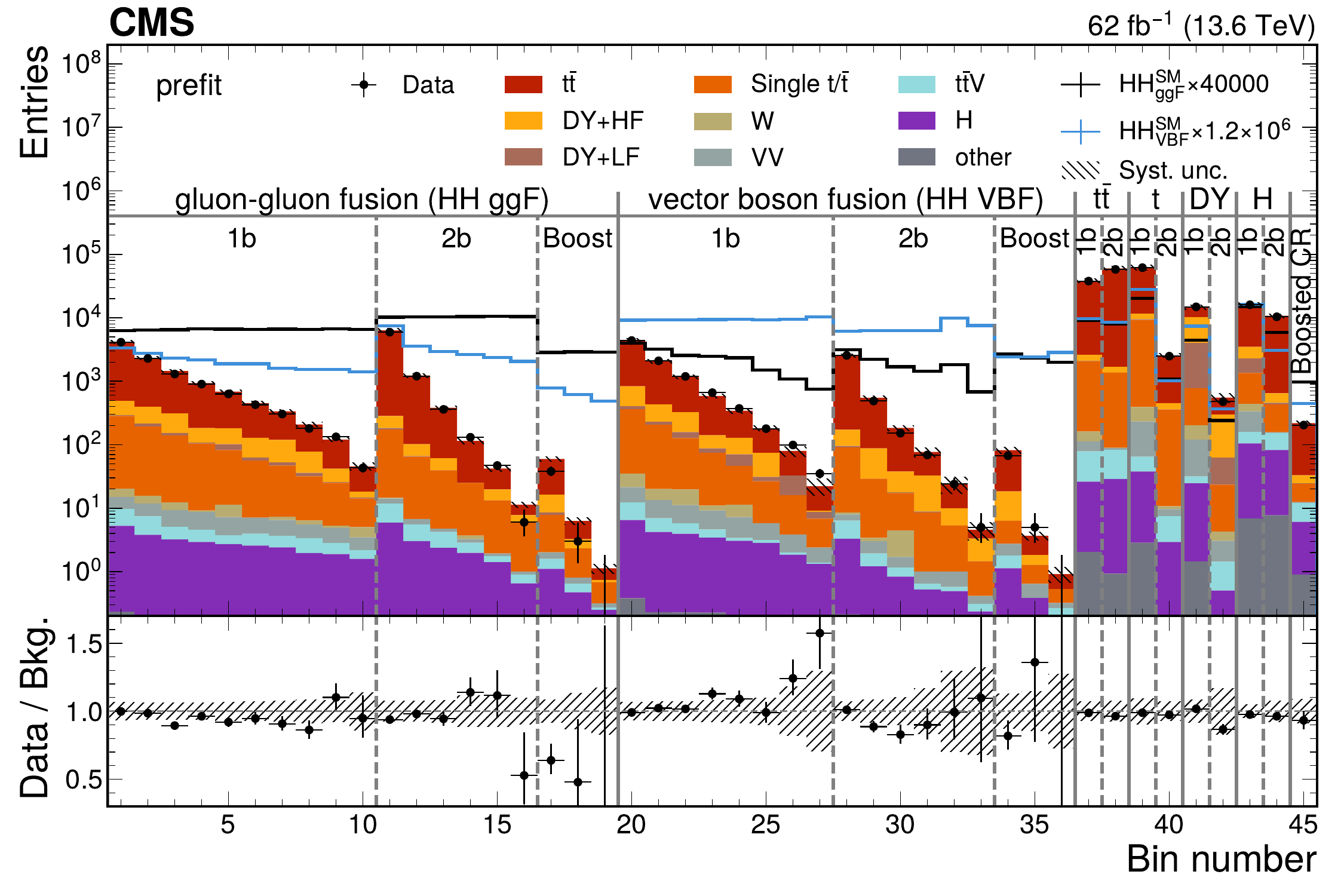}
  \includegraphics[width=0.9\textwidth]{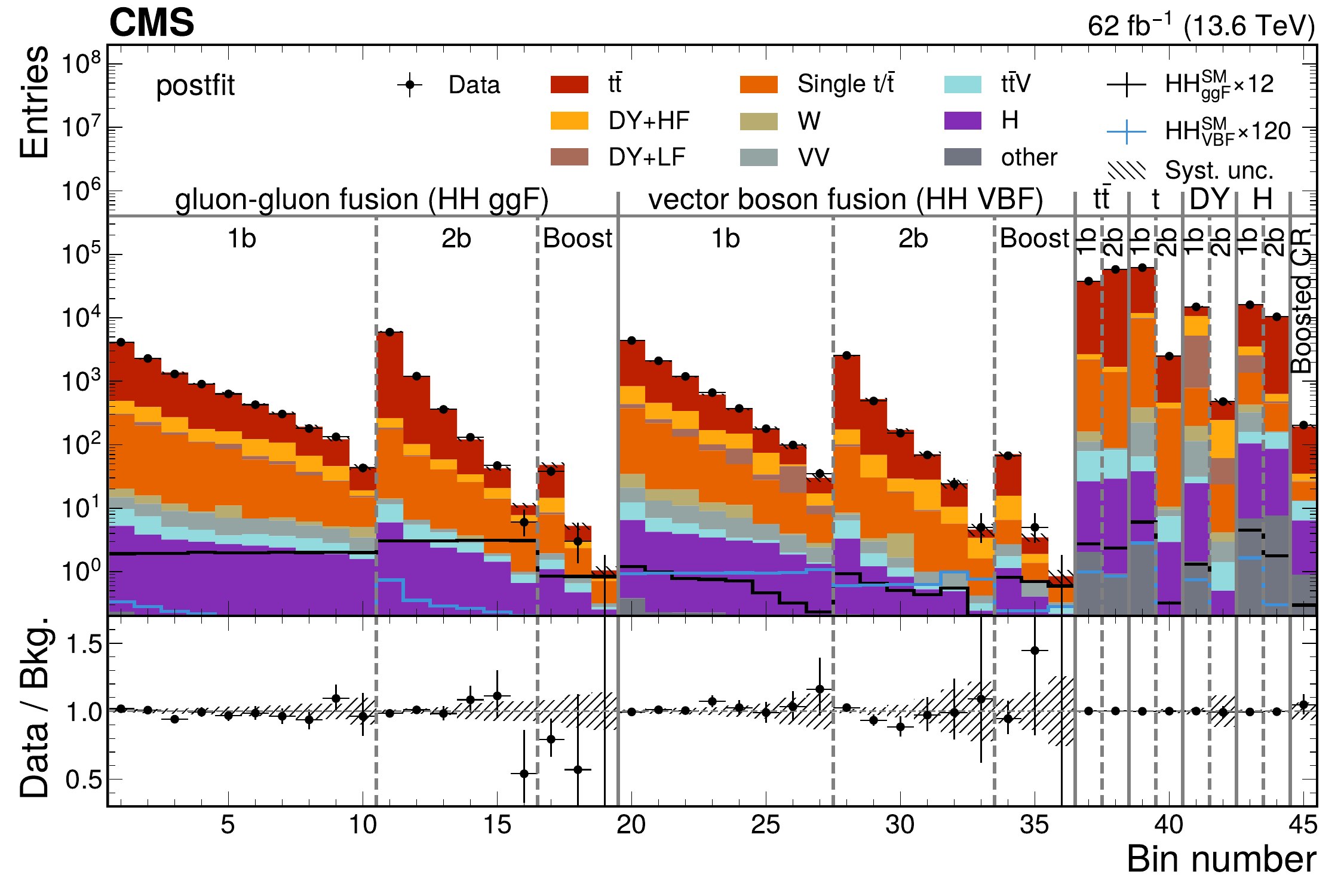}
  \caption{%
    Observed (points) and expected (filled histograms) yields in each
    discriminant (NN score or category yield) bin before (upper) and
    after (lower) the fit to data.
    The \HH signal distributions in the \ggF and \VBF
    production channels are overlaid (solid lines), scaled to the
    total background yield (top) or the observed upper limit for \ggF
    and the observed upper limit times 10 for \VBF (bottom).
    The uncertainty bands include the total uncertainty of the fit
    model.
    The lower panels show the ratio of the data to the expected
    background yields.
  }%
  \label{fig:classifiers}%
\end{figure}

A simultaneous binned profile likelihood fit of the signal and
background templates to the observed discriminant distributions in all
categories is performed, using the CMS statistical analysis tool
{\scshape Combine}~\cite{CMS:2024onh}.
Each template includes the events of all data-taking periods.
The rates of the \ttbar and \DY background processes described in Section~\ref{sec:strategy:bkgmodelling} are free parameters in the
fit, and the systematic uncertainties described in
Section~\ref{sec:systematics} are taken into account via
nuisance parameters in the likelihood function.
Log-normal distributions are used to model the effects of nuisance
parameters on the overall rate.
Shape uncertainties are implemented through an interpolation between
alternative templates~\cite{CMS:2024onh,Conway:2011in}, obtained by
propagating the effect of the uncertainties to the input variables of
the NNs and then evaluating the NNs using the varied inputs.
Uncertainties that are decorrelated between different data-taking periods are implemented by varying only the fractional contributions corresponding to the data-taking periods in each template.
The observed yields in each bin of the final discriminant
distributions in all channels and categories are shown in
Fig.~\ref{fig:classifiers}.

The goodness-of-fit is quantified using a $p$-value that takes into account
the postfit uncertainty model~\cite{CMS:2024onh} and amounts to $p=\resultGoF$,
indicating good description of the data by the fit model.
The best fit value of the \ttbar  background normalisation parameter is
$\normTTres^{+\normTTresUp}_{-\normTTresDn}$ ($\normTTboost^{+\normTTboostUp}_{-\normTTboostDn}$ in the boosted
categories), consistent with a dedicated \ttbar cross section
measurement in a similar phase space region~\cite{CMS:2023qyl}.

No significant excess of events over the background-only hypothesis is
observed.
Upper limits on the cross section of non-resonant Higgs boson pair production are
set at 95\% \CL under the background-only hypothesis, using the \CLs
criterion~\cite{Junk:1999kv,Read:2002hq} and assuming asymptotic
distributions~\cite{Cowan:2010js} of the profile likelihood ratio test
statistics defined in Ref.~\cite{CMS:2024onh}.
The observed upper limit on the inclusive \HH production
cross section normalised to the value expected in the SM (\HH signal
strength modifier) is \resultLimitHHobs, with a median expected limit
of \resultLimitHHexp.
The observed limit is lower than the expected limit, which can be
attributed to a downward fluctuation in the observed event counts
relative to the background prediction in several bins, most notably in
the most sensitive \SRggF bin (bin 16 in Fig.~\ref{fig:classifiers}).

\begin{figure}[htp]
  \centering
  \includegraphics[width=0.75\textwidth]{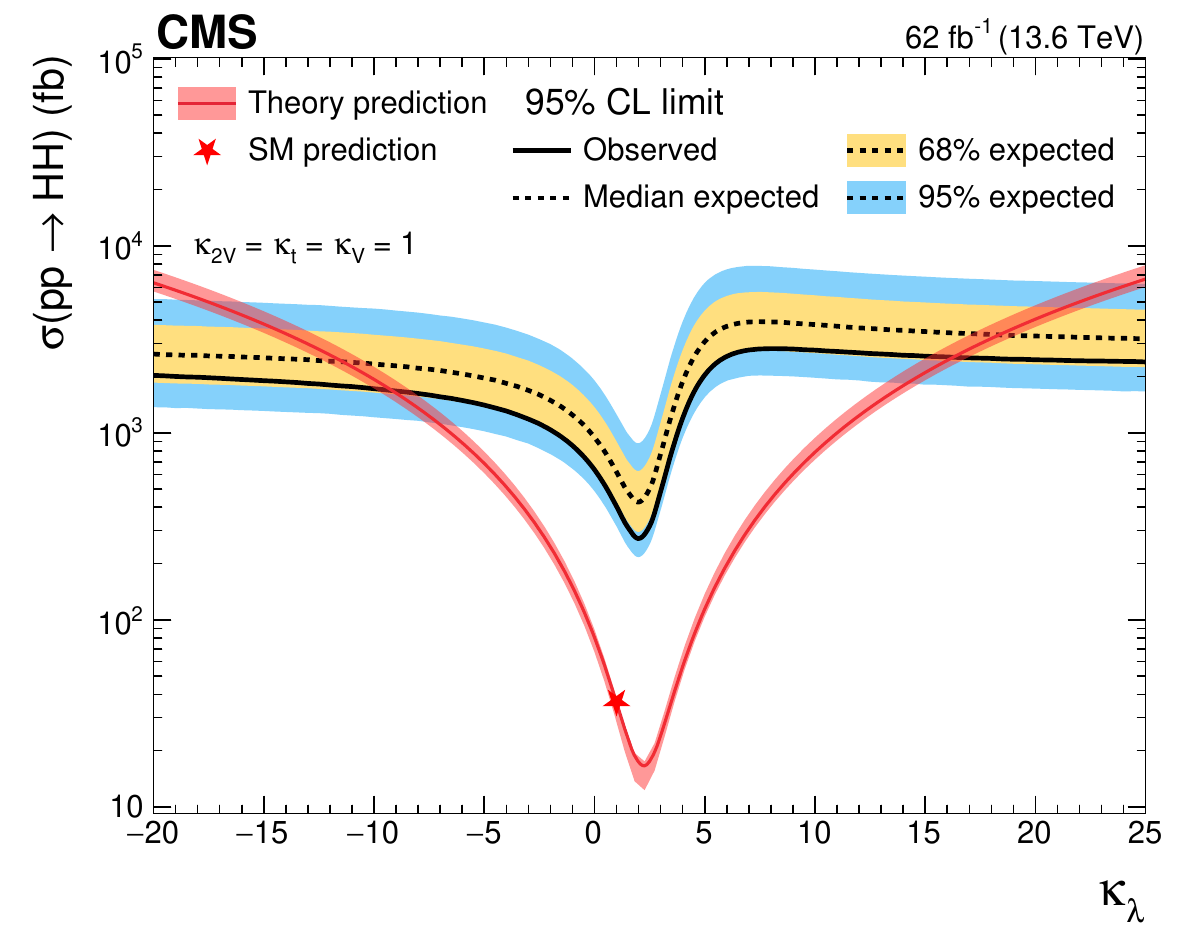}
  \includegraphics[width=0.75\textwidth]{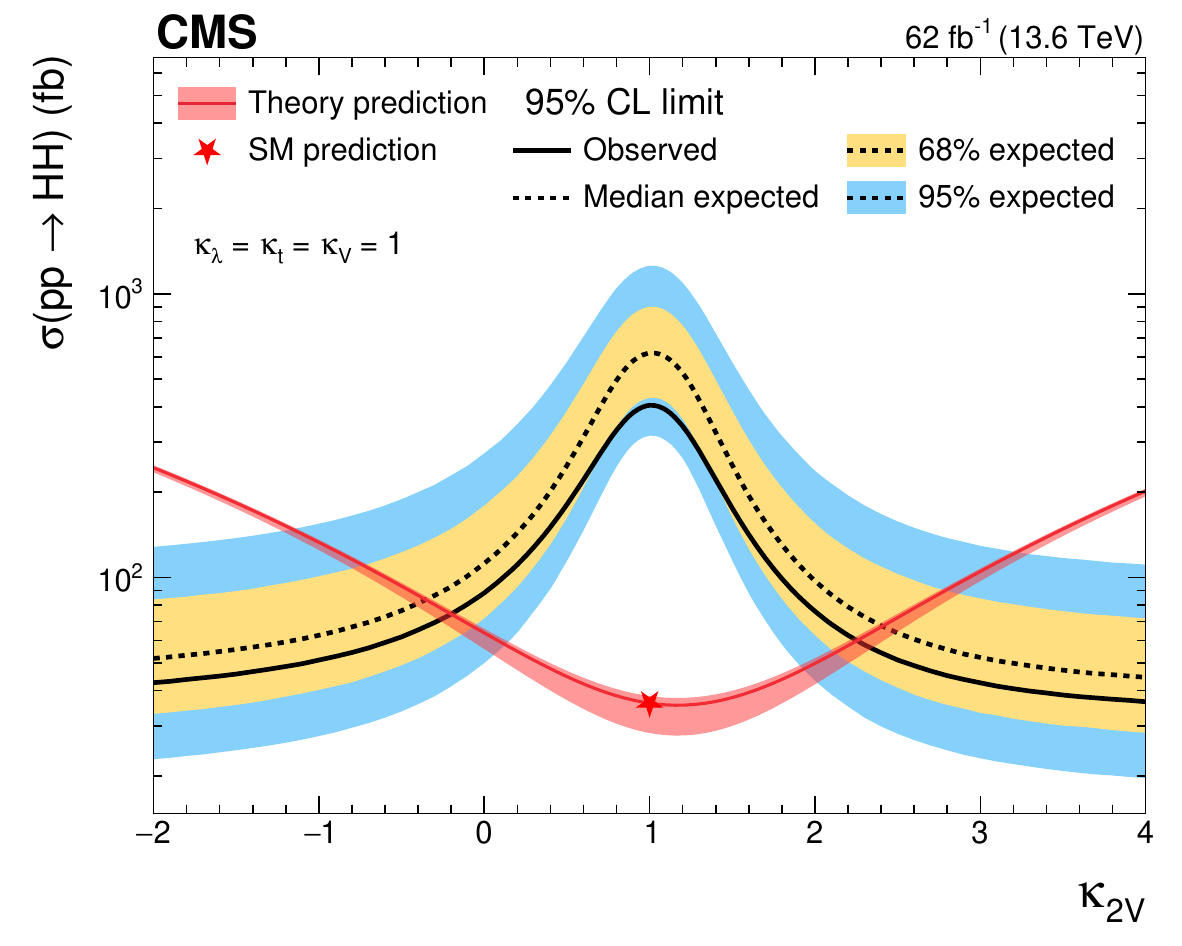}
  \caption{
    Observed (solid black line) and median expected (dashed black line) upper limits at
    the 95\% \CL on the inclusive \HH production cross section as a
    function of \kappaL (upper) and \kappaVV (lower);
    in both cases, all respective other couplings are fixed to the SM
    prediction.
    The yellow (blue) bands show the 68\% (95\%) confidence level
    intervals of the expected limit.
    The predicted cross section is overlaid (red curve), and the SM
    prediction is indicated (red star).
  }
  \label{fig:kappascans}
\end{figure}

The limit on the \HH production cross section is also shown as a
function of \kappaL and as a function of \kappaVV in
Fig.~\ref{fig:kappascans}, where in each case all other couplings are
assumed to conform to the SM prediction.
The strong dependence of the limit on \kappaVV is a result of the
increased selection efficiency and acceptance to \VBF signal events in
the boosted regime for values of \kappaVV different from 1.
At $\kappaVV=1$, the signal sensitivity is equally driven by the
\catbb and the \catboosted categories, while for \kappaVV different
from 1, the sensitivity remains the same in the \catbb categories but
strongly improves in the \catboosted categories.
For the results in Fig.~\ref{fig:kappascans}, the uncertainty in the
inclusive \HH cross section prediction is not included in the fits but
is represented as an uncertainty band on the theory prediction shown.

\begin{figure}[htp]
  \centering
  \includegraphics[width=0.48\textwidth]{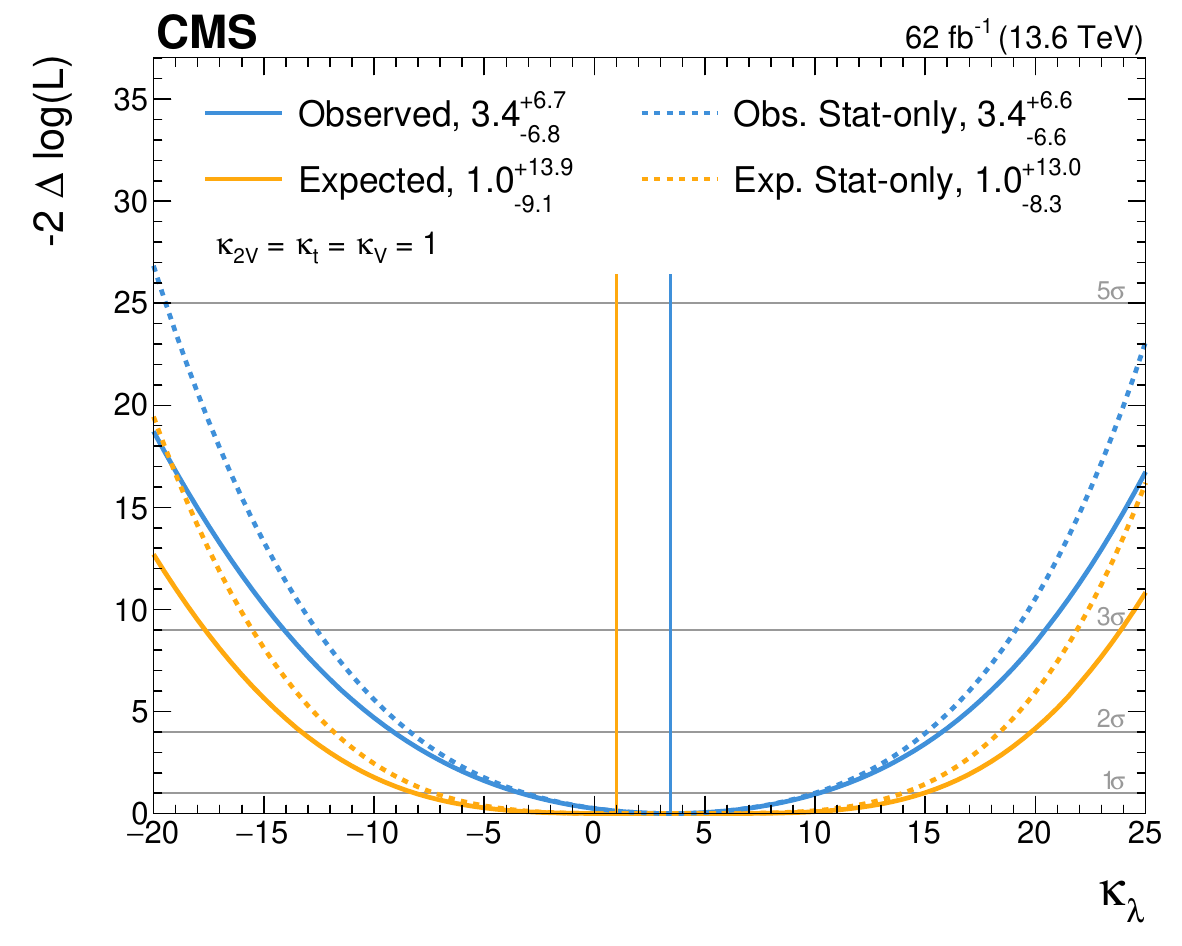}
  \includegraphics[width=0.48\textwidth]{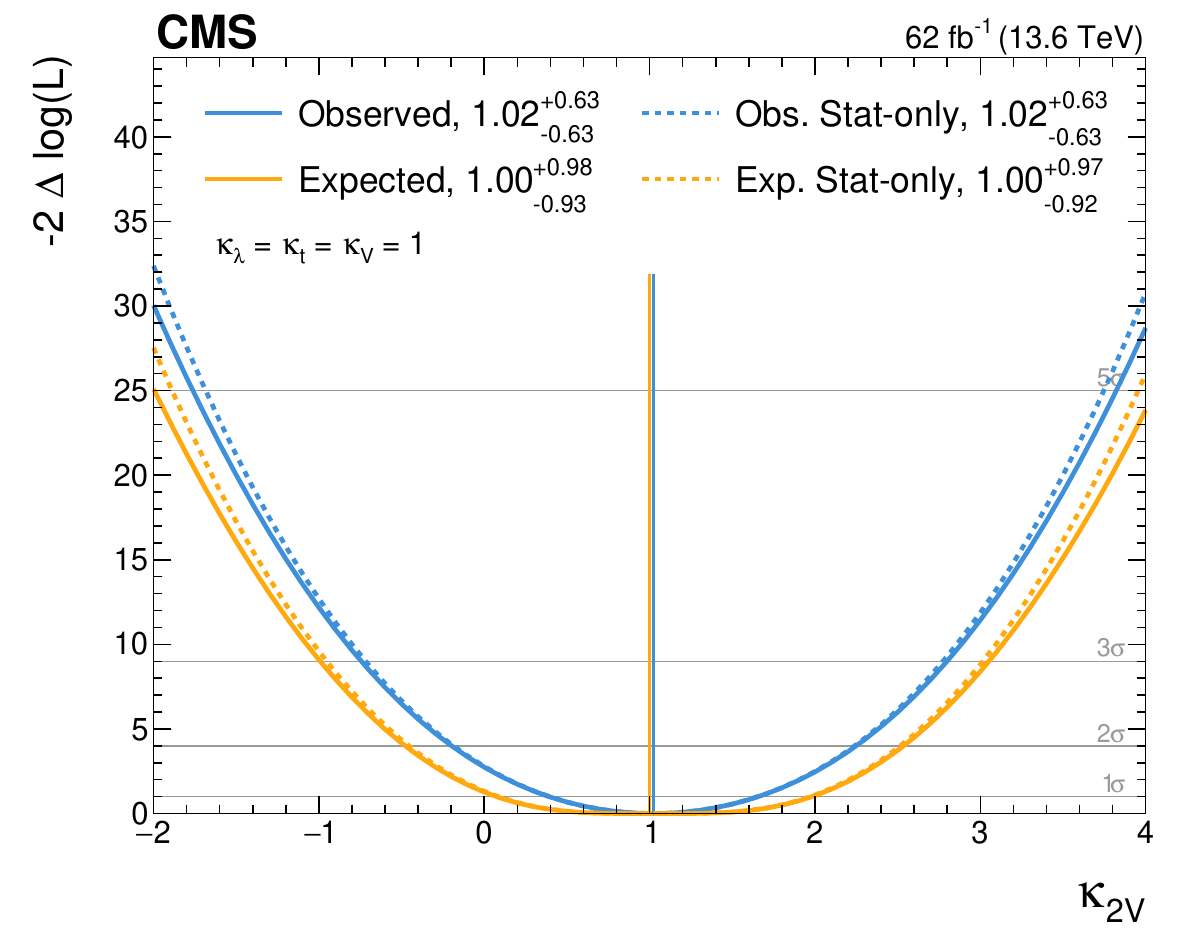}
  \caption{
    Observed (blue) and expected (orange) negative
    log-likelihood values as a function of \kappaL (left) and \kappaVV
    (right), assuming all other couplings conform to the SM
    prediction.
    The solid lines include the full uncertainty model, and the dashed
    lines only include statistical uncertainties, which include the
    uncertainty components due to the \ttbar and \DY background
    normalisations.
    The vertical lines indicate the best fit value.
  }
  \label{fig:LLscans}
\end{figure}

\begin{figure}[htp]
  \centering
  \includegraphics[width=0.75\textwidth]{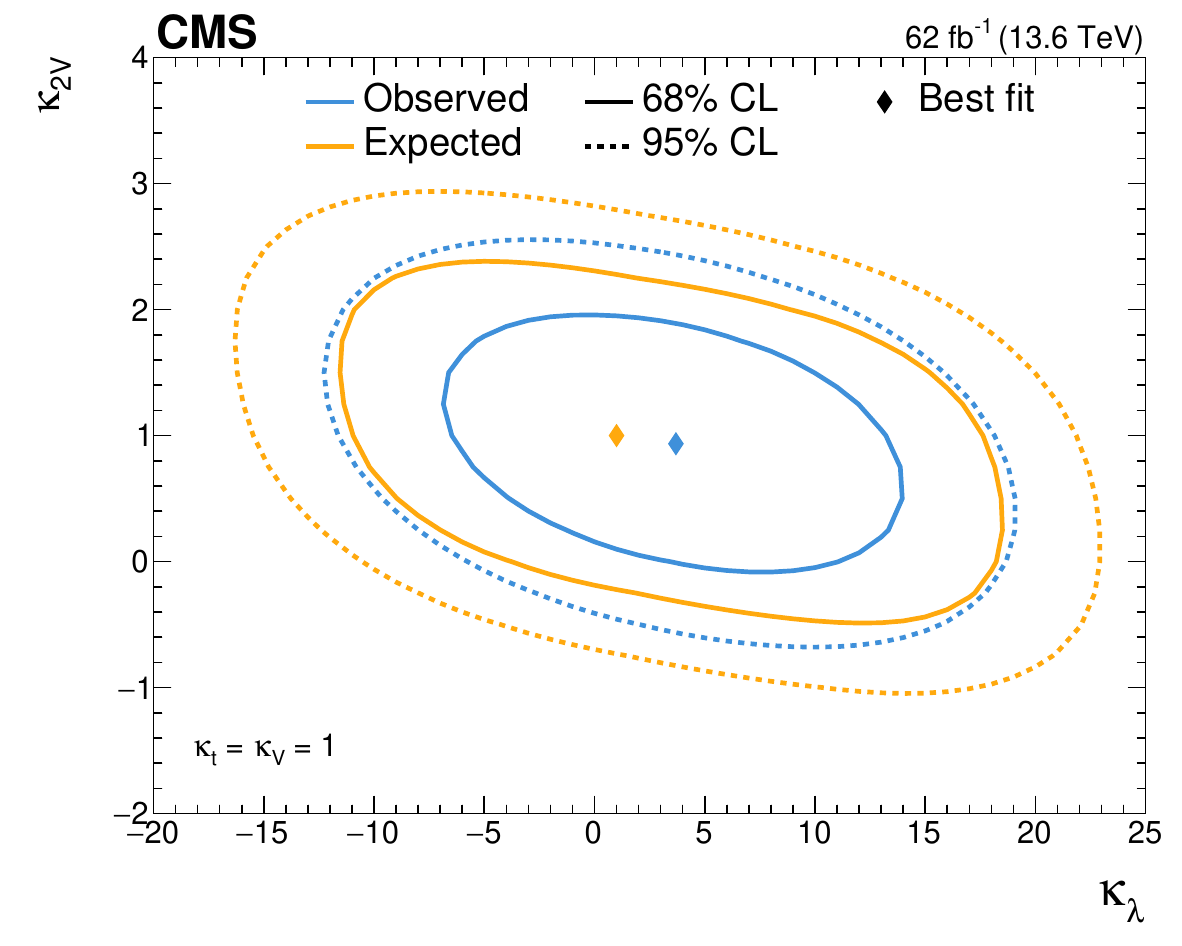}
  \caption{
    Observed (blue) and expected (orange) negative log-likelihood
    contours as a function of \kappaL and \kappaVV, assuming all other
    couplings conform to the SM prediction.
    Shown are the best fit point (marker) and the 68\% (solid lines)
    and 95\% (dashed lines) \CL contours.
  }
  \label{fig:LLscans2D}
\end{figure}

In Fig.~\ref{fig:LLscans}, the observed and expected profile
likelihood ratio values are presented as functions of \kappaL and
\kappaVV, where again all other couplings are assumed as in the SM.
The best fit value of \kappaL is \resultKappaL, consistent with the
downward fluctuation of data in the \SRggF and the minimal \HH
production cross section near $\kappaL=2.5$.
Values of \kappaL outside $[\resultKappaLMinObs,\resultKappaLMaxObs]$
($[\resultKappaLMinExp,\resultKappaLMaxExp]$ expected) and of \kappaVV
outside $[\resultKappaVVMinObs,\resultKappaVVMaxObs]$
($[\resultKappaVVMinExp,\resultKappaVVMaxExp]$ expected) are excluded
at the 95\% \CL.
The likelihood contours from a simultaneous fit of \kappaL and
\kappaVV are shown in Fig.~\ref{fig:LLscans2D}.
The results with statistical uncertainty only are obtained by
first performing the nominal fit including systematic uncertainties,
and then evaluating the likelihood with the values of the nuisance
parameters used to describe the systematic uncertainties fixed to
their best fit values.
The background normalisation parameters are not fixed and thus
contribute to the statistical uncertainties.

By far the largest source of uncertainty in the analysis arises from the limited statistical precision due to the small signal rate.
The impact of the statistical uncertainties compared to the systematic effects is shown in Fig.~\ref{fig:LLscans}.

\begin{figure}[!hpt]
  \centering
  \includegraphics[width=0.9\textwidth]{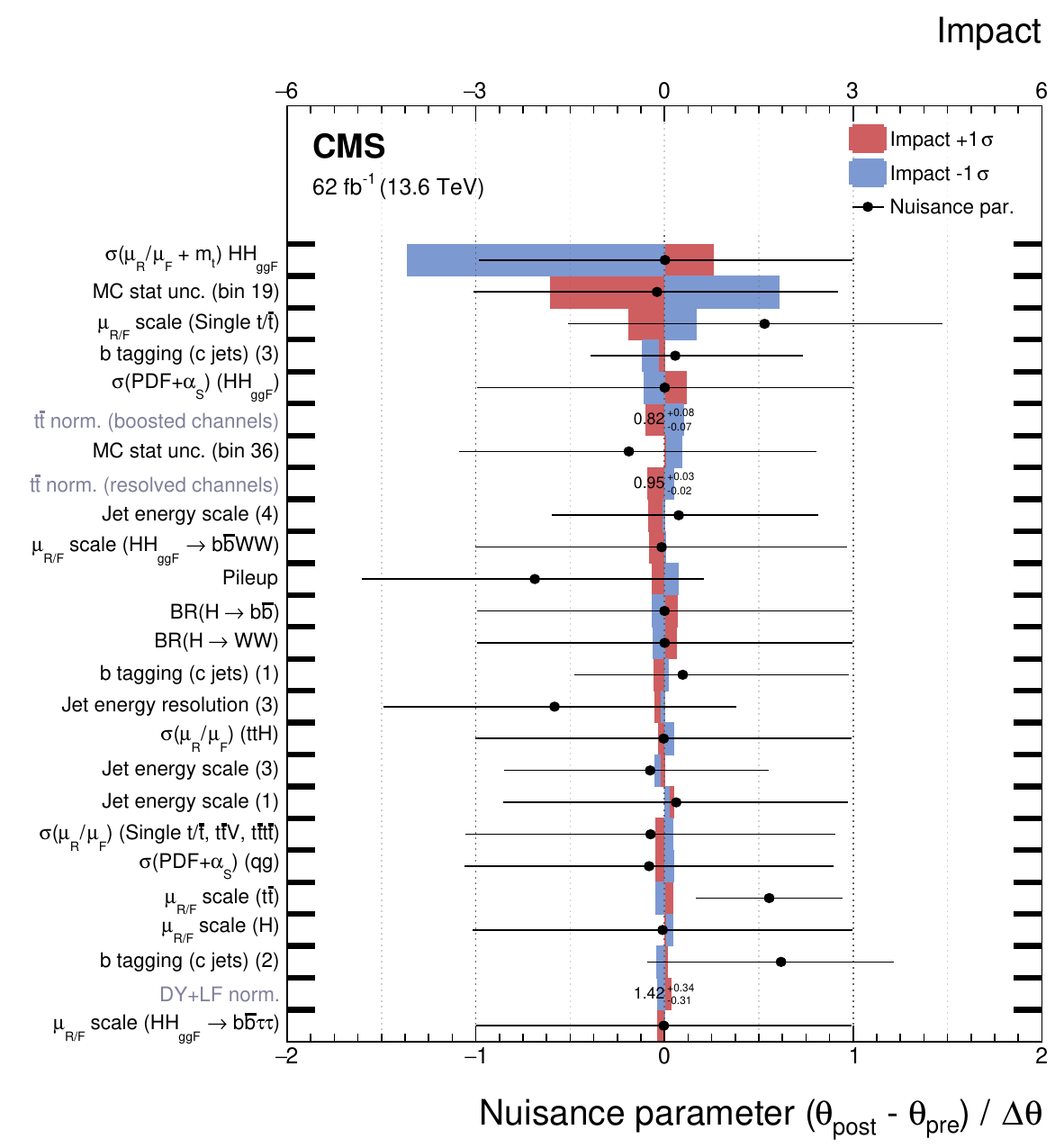}
  \caption{%
    Best fit values of the
    background normalisation and nuisance parameters (black markers).
    The nuisance parameter values are shown
    as the difference of their best fit values, $\theta_{\text{post}}$, and
    prefit values, $\theta_{\text{pre}}$, relative to the prefit uncertainties
    $\Delta\theta$.
    The impact (coloured areas) of the nuisance
    parameters on the \HH signal strength is computed as the
    difference of the nominal best fit value of the signal strength
    and the best fit
    value obtained when fixing the nuisance parameter under scrutiny
    to its best fit value $\theta_{\text{post}}$ plus/minus its postfit
    uncertainty.
    The nuisance parameters are ordered
    by their impact, and only the 25 highest-ranked parameters are
    shown.
    The number in parentheses for the jet energy scale and \PQb tagging uncertainties correspond to a numbering of the data-taking period to which they are associated.
    The MC stat unc. refers to the systematic uncertainty due to the
    limited number of simulated events; in this case, the number in parentheses refers to the bin numbers shown in Fig.~\ref{fig:classifiers}.
  }
  \label{fig:impacts}
\end{figure}

The best fit values of the nuisance parameters used to describe the systematic uncertainties are shown in Fig.~\ref{fig:impacts}; they are well within one
standard deviation of the prior uncertainty.
The impact of each normalisation and nuisance parameter on the result is evaluated as
the difference of the nominal best fit value of the \HH signal
strength modifier and the best fit value obtained when fixing the
parameter under scrutiny to its best fit value plus/minus its postfit
uncertainty.
The best fit values and impacts of the 25 parameters with the highest
impact are shown in Fig.~\ref{fig:impacts}; the nuisance parameters
with the largest impact are related to the uncertainty in the inclusive \HH
cross section prediction (which enters the measured signal strength
modifier in the denominator) and the limited size of the simulated
event samples.

The results are combined with the results of the previous search with
138\fbinv of Run 2 data collected at $\sqrt{s}=13\TeV$ in final states
with one or two leptons~\cite{CMS:2024rgy}.
Systematic uncertainties of theoretical sources, such as cross section
calculations, are treated as correlated in the combination, while
experimental uncertainties, which are dominated by the uncertainties
in auxiliary measurements in statistically independent data sets, are
treated as uncorrelated.
An observed upper limit on the \HH signal strength modifier of \resultCombinationLimitHHobs at 95\% \CL is obtained, with an expectation of \resultCombinationLimitHHexp.

\section{Summary}\label{sec:summary}
A search has been presented for Higgs boson pair
production in the \bbWW decay channel with two leptons in the final
state, conducted with \lumitot of proton-proton collision data collected at $\sqrt{s}=13.6\TeV$.

The data are consistent with standard model predictions.
An upper limit is set on the Higgs boson pair production cross section
of \resultLimitHHobs times the standard model prediction at the 95\%
confidence level, with an expectation of \resultLimitHHexp.
Compared to a previous search by the CMS Collaboration with 138\fbinv
of Run 2 data collected at $\sqrt{s}=13\TeV$ in final states with one or two leptons, the sensitivity
is significantly improved through a refined classification strategy,
additional triggers, and an enhanced \PQb tagging algorithm.
This yields a comparable overall expected sensitivity despite the
smaller analysed data set and restriction to the two-lepton channel,
with a 30\% sensitivity increase in the two-lepton final state alone.
The cross section limit is further used to constrain the trilinear
self-coupling of the Higgs boson and the
quartic coupling between two Higgs bosons and two vector bosons to
$[\resultKappaLMinObs,\resultKappaLMaxObs]$
($[\resultKappaLMinExp,\resultKappaLMaxExp]$ expected) and $[\resultKappaVVMinObs,\resultKappaVVMaxObs]$
($[\resultKappaVVMinExp,\resultKappaVVMaxExp]$ expected) times the
standard model expectation, respectively, at 95\% confidence level.

The presented search is the first result on Higgs boson pair
production in the \bbWW decay channel with $\sqrt{s}=13.6\TeV$ data.
Combined with the results of the previous search at $\sqrt{s}=13\TeV$
in final states with one or two leptons, an observed upper limit is
set on the Higgs boson pair production cross section of
\resultCombinationLimitHHobs times the standard model prediction at
the 95\% confidence level, with an expectation of
\resultCombinationLimitHHexp.

\begin{acknowledgments}
We congratulate our colleagues in the CERN accelerator departments for the excellent performance of the LHC and thank the technical and administrative staffs at CERN and at other CMS institutes for their contributions to the success of the CMS effort. In addition, we gratefully acknowledge the computing centres and personnel of the Worldwide LHC Computing Grid and other centres for delivering so effectively the computing infrastructure essential to our analyses. Finally, we acknowledge the enduring support for the construction and operation of the LHC, the CMS detector, and the supporting computing infrastructure provided by the following funding agencies: SC (Armenia), BMBWF and FWF (Austria); FNRS and FWO (Belgium); CNPq, CAPES, FAPERJ, FAPERGS, and FAPESP (Brazil); MES and BNSF (Bulgaria); CERN; CAS, MoST, and NSFC (China); MINCIENCIAS (Colombia); MSES and CSF (Croatia); RIF (Cyprus); SENESCYT (Ecuador); ERC PRG and PSG, TARISTU24-TK10 and MoER TK202 (Estonia); Academy of Finland, MEC, and HIP (Finland); CEA and CNRS/IN2P3 (France); SRNSF (Georgia); BMFTR, DFG, and HGF (Germany); GSRI (Greece); NKFIH (Hungary); DAE and DST (India); IPM (Iran); SFI (Ireland); INFN (Italy); MSIT and NRF (Republic of Korea); MES (Latvia); LMTLT (Lithuania); MOE and UM (Malaysia); BUAP, CINVESTAV, CONACYT, LNS, SEP, and UASLP-FAI (Mexico); MOS (Montenegro); MBIE (New Zealand); PAEC (Pakistan); MSHE, NSC, and NAWA (Poland); FCT (Portugal); MESTD (Serbia); MICIU/AEI and PCTI (Spain); MOSTR (Sri Lanka); Swiss Funding Agencies (Switzerland); MST (Taipei); MHESI (Thailand); TUBITAK and TENMAK (T\"{u}rkiye); NASU (Ukraine); STFC (United Kingdom); DOE and NSF (USA).

\hyphenation{Rachada-pisek} Individuals have received support from the Marie-Curie programme and the European Research Council and Horizon 2020 Grant, contract Nos.\ 675440, 724704, 752730, 758316, 765710, 824093, 101115353, 101002207, 101001205, and COST Action CA16108 (European Union); the Leventis Foundation; the Alfred P.\ Sloan Foundation; the Alexander von Humboldt Foundation; the Science Committee, project no. 22rl-037 (Armenia); the Fonds pour la Formation \`a la Recherche dans l'Industrie et dans l'Agriculture (FRIA) and Fonds voor Wetenschappelijk Onderzoek contract No. 1228724N (Belgium); the Beijing Municipal Science \& Technology Commission, No. Z191100007219010, the Fundamental Research Funds for the Central Universities, the Ministry of Science and Technology of China under Grant No. 2023YFA1605804, the Natural Science Foundation of China under Grant No. 12535004, and USTC Research Funds of the Double First-Class Initiative No.\ YD2030002017 (China); the Ministry of Education, Youth and Sports (MEYS) of the Czech Republic; the Shota Rustaveli National Science Foundation, grant FR-22-985 (Georgia); the Deutsche Forschungsgemeinschaft (DFG), among others, under Germany's Excellence Strategy -- EXC 2121 ``Quantum Universe" -- 390833306, and under project number 400140256 - GRK2497; the Hellenic Foundation for Research and Innovation (HFRI), Project Number 2288 (Greece); the Hungarian Academy of Sciences, the New National Excellence Program - \'UNKP, the NKFIH research grants K 131991, K 133046, K 138136, K 143460, K 143477, K 146913, K 146914, K 147048, 2020-2.2.1-ED-2021-00181, TKP2021-NKTA-64, and 2025-1.1.5-NEMZ\_KI-2025-00004 (Hungary); the Council of Science and Industrial Research, India; ICSC -- National Research Centre for High Performance Computing, Big Data and Quantum Computing, FAIR -- Future Artificial Intelligence Research, and CUP I53D23001070006 (Mission 4 Component 1), funded by the NextGenerationEU program, the Italian Ministry of University and Research (MUR) under Bando PRIN 2022 -- CUP I53C24002390006, PRIN PRIMULA 2022RBYK7T (Italy); the Latvian Council of Science; the Ministry of Education and Science, project no. 2022/WK/14, and the National Science Centre, contracts Opus 2021/41/B/ST2/01369, 2021/43/B/ST2/01552, 2023/49/B/ST2/03273, and the NAWA contract BPN/PPO/2021/1/00011 (Poland); the Funda\c{c}\~ao para a Ci\^encia e a Tecnologia (Portugal); the National Priorities Research Program by Qatar National Research Fund; MICIU/AEI/10.13039/501100011033, ERDF/EU, ``European Union NextGenerationEU/PRTR", projects PID2022-142604OB-C21, PID2022-139519OB-C21, PID2023-147706NB-I00, PID2023-148896NB-I00, PID2023-146983NB-I00, PID2023-147115NB-I00, PID2023-148418NB-C41, PID2023-148418NB-C42, PID2023-148418NB-C43, PID2023-148418NB-C44, PID2024-158190NB-C22, RYC2021-033305-I, RYC2024-048719-I, CNS2023-144781, CNS2024-154769 and Plan de Ciencia, Tecnolog{\'i}a e Innovaci{\'o}n de Asturias, Spain; the Chulalongkorn Academic into Its 2nd Century Project Advancement Project, the National Science, Research and Innovation Fund program IND\_FF\_68\_369\_2300\_097, and the Program Management Unit for Human Resources \& Institutional Development, Research and Innovation, grant B39G680009 (Thailand); the Eric \& Wendy Schmidt Fund for Strategic Innovation through the CERN Next Generation Triggers project under grant agreement number SIF-2023-004; the Kavli Foundation; the Nvidia Corporation; the SuperMicro Corporation; the Welch Foundation, contract C-1845; and the Weston Havens Foundation (USA).
\end{acknowledgments}

\bibliography{auto_generated}
\cleardoublepage \appendix\section{The CMS Collaboration \label{app:collab}}\begin{sloppypar}\hyphenpenalty=5000\widowpenalty=500\clubpenalty=5000\cmsinstitute{Yerevan Physics Institute, Yerevan, Armenia}
{\tolerance=6000
A.~Gevorgyan\cmsorcid{0000-0003-2751-9489}, A.~Hayrapetyan, V.~Makarenko\cmsorcid{0000-0002-8406-8605}, A.~Tumasyan\cmsAuthorMark{1}\cmsorcid{0009-0000-0684-6742}
\par}
\cmsinstitute{Marietta Blau Institute for Particle Physics, Vienna, Austria}
{\tolerance=6000
W.~Adam\cmsorcid{0000-0001-9099-4341}, L.~Benato\cmsorcid{0000-0001-5135-7489}, T.~Bergauer\cmsorcid{0000-0002-5786-0293}, M.~Dragicevic\cmsorcid{0000-0003-1967-6783}, S.~Gundacker\cmsorcid{0000-0003-2087-3266}, A.K.~Guven\cmsorcid{0009-0004-5670-5138}, P.S.~Hussain\cmsorcid{0000-0002-4825-5278}, M.~Jeitler\cmsAuthorMark{2}\cmsorcid{0000-0002-5141-9560}, N.~Krammer\cmsorcid{0000-0002-0548-0985}, A.~Li\cmsorcid{0000-0002-4547-116X}, D.~Liko\cmsorcid{0000-0002-3380-473X}, M.~Matthewman, J.~Schieck\cmsAuthorMark{2}\cmsorcid{0000-0002-1058-8093}, R.~Sch\"{o}fbeck\cmsAuthorMark{2}\cmsorcid{0000-0002-2332-8784}, M.~Shooshtari\cmsorcid{0009-0004-8882-4887}, M.~Sonawane\cmsorcid{0000-0003-0510-7010}, N.~Van~Den~Bossche\cmsorcid{0000-0003-2973-4991}, W.~Waltenberger\cmsorcid{0000-0002-6215-7228}, C.-E.~Wulz\cmsAuthorMark{2}\cmsorcid{0000-0001-9226-5812}
% ...
\par}
\vskip\cmsinstskip
\dag:~Deceased\\
$^{1}$Also at Yerevan State University, Yerevan, Armenia\\
$^{2}$Also at TU Wien, Vienna, Austria\\
% ...
\end{sloppypar}
\end{document}